\documentstyle[12pt]{article}
\def\llsymbol#1{\@llsymbol{\@nameuse{c@#1}}}
\def\@llsymbol#1{\ifcase#1\or {}\or {'}\or {''}\or {'''}\or 
   {''''}\or {'''''}\or  \else\@ctrerr\fi\relaz}

\newcounter{contador}
\newcommand{\letra}{
   \stepcounter{equation}
   \setcounter{contador}{\value{equation}}
   \setcounter{equation}{0}
   \renewcommand{\theequation}{\thecontador\alph{equation}}}
\newcommand{\antiletra}{
   \renewcommand{\theequation}{\arabic{equation}}
   \setcounter{equation}{\value{contador}}}

\begin{document}
\title{ Integral relations and new solutions to the double-confluent 
Heun equation}
\author{Bartolomeu D B Figueiredo (e-mail: barto@cbpf.br)
\\
{\small Centro Brasileiro de Pesquisas F\'{\i}sicas---CBPF, 
 Rua Xavier Sigaud, 150, }\\
{\small 22290-180, Rio de Janeiro, Brazil}}
\maketitle
\begin{center}{\small Draft version}\end{center} 
%
%                               ABSTRACT
%               
\begin{abstract}
\noindent
Integral relations and transformation rules are used to obtain, out of
an asymptotic solution, a new group of four pairs of solutions to the
double-confluent Heun equation. Each pair presents the same series
coefficients but has solutions convergent in different regions of the
complex plane.  Integral relations are also established between
solutions given by series of Coulomb wave functions. The Whittaker-Hill
equation and another equation are studied as particular cases of both
the double-confluent and the single-confluent Heun equations.  Finally,
applications for the Schr\"{o}dinger equation with certain potentials
are discussed, mainly for quasi-exactly solvable potentials which lead
to the above special equations.

%
%\newpage
%
\end{abstract}
%
%
%				SECTION I
%
\section*{1. Preliminary remarks}
Here we deal with two groups of series solutions to the
double-confluent Heun equation (DCHE). The first group, constituted by
four pairs of solutions, is generated from an asymptotic expansion by
means of integral relations and transformation rules, and the second
group is given by pairs of solutions in series of Coulomb wave
functions, already derived in \cite{eu}.  For the latter, we show that
an integral relation also exists between the members of each pair, and
we provide additional properties for the solutions. After this, we
analyse two differential equations which are special cases of both the
DCHE and the single-confluent Heun equation and, finally, we use some
results to solve the Schr\"{o}dinger equation for certain potentials.
Before proceeding, we set down some conventions, present the procedures
used to obtain the solutions and outline the structure of the paper.

For the DCHE we adopt the form 
\begin{eqnarray}
\label{dche}
z^2\frac{d^{2}U}{dz^{2}}+(B_{1}+B_{2}z)\frac{dU}{dz}+ 
\left(B_{3}-2\eta\omega z+\omega^{2}z^2\right)U=0,\ \ (B_{1}\neq 0,\
\ \omega\neq 0)
\end{eqnarray}
where $z=0$ and $z=\infty$ are irregular singularities.  $B_{1}=0$
and/or $\omega=0$ are excluded because, in these cases, the equation
reduces to a confluent hypergeometric equation or to an equation with
constant coefficients. Thus, if $B_{1}=0$ and $\omega\neq 0$, the
substitutions
\letra
\begin{eqnarray}
 y=-2i\omega z, \ \  U(z)=e^{-y/2}y^{\alpha}f(y), \ \ 
\alpha^2-(1-B_{2})\alpha+B_{3}=0
\end{eqnarray}
give the confluent hypergeometric equation
\begin{eqnarray}
y\frac{d^2f}{dy^2}+[(2\alpha+B_{2})-y]\frac{df}{dy}-
\left(i\eta+\alpha+\frac{B_{2}}{2}\right)f=0
\end{eqnarray}
and, if $B_{1}\neq 0$ and $\omega= 0$, 
\antiletra
\letra
\begin{eqnarray}
&&y=B_{1}/z, \ \  U(z)=y^{\beta}g(y), \ \ \beta^2-(B_{2}-1)\beta+B_{3}=0
\Rightarrow\\
&&y\frac{d^2g}{dy^2}+[(2\beta+2-B_{2})-y]\frac{dg}{dy}-\beta g=0.
\end{eqnarray}
If $B_{1}=\omega=0$, we find an equation with constant coefficients by
taking $z=\exp{y}$.

Equation (\ref{dche}) can be obtained from the generalized spheroidal
wave equation (GSWE) \cite{wilson}, which is also known as
single-confluent Heun equation or, simply, confluent Heun equation
\cite{ronveaux}.  Actually the GSWE in Leaver's form reads
\cite{leaver1}
\antiletra
\begin{eqnarray}
\label{gswe1}
z(z-z_{0})\frac{d^{2}U}{dz^{2}}+(B_{1}+B_{2}z)\frac{dU}{dz}+ 
\left[B_{3}-2\eta \omega(z-z_{0})+\omega^{2}z(z-z_{0})\right]U=0,
\ (\omega \neq 0)
\end{eqnarray}
where $z_{0}$, $B_{i}$, $\eta$ and $\omega$ are constants, $z=0$ and
$z=z_{0}$ are regular singular points while $z=\infty$ is an irregular
singularity. For $z_{0}=0$ the preceding equation gives the DCHE and,
when a limiting solution also exits, this one can be taken as a
starting-point to generate a group of solutions to the DCHE. This
procedure was applied by Leaver to find expansions in series of Coulomb
wave functions for the DCHE, although these can as well be found
otherwise \cite{schmidt}. Note that, although the DCHE was called
`confluent GSWE' in \cite{eu,leaver1}, it is not derived from the GSWE
by a process of confluence (see \cite{slavyanov}, chapter 4).

The transformation rules aforementioned result from variable
substitutions that convert the DCHE into another version of itself.
Thus, for a given solution $S(z)$,
\begin{eqnarray}
\label{sol1}
S(z):=U(B_{1},B_{2},B_{3};\omega,\eta;z),
\end{eqnarray} 
(`$:=$' means `equal by definition') we shall generate
new solutions by using the following rules [1,3,6-8] ---
$r_{1},\ r_{2},\ r_{3}$ ---
\begin{eqnarray}
&&r_{1}S(z)=e^{i\omega z+{B_{1}}/({2z})}z^{-i\eta-{B_{2}}/{2}}
U(B_{1}^{'},B_{2}^{'},B_{3}^{'};\omega^{'},\eta^{'};\vartheta),\\
\mbox{}\nonumber \\
&&r_{2}S(z)=e^{{B_{1}}/{z}}z^{2-B_{2}}U(-B_{1},4-B_{2},
B_{3}+2-B_{2}; \omega,\eta;z),\\
\mbox{}\nonumber \\
&&r_{3}S(z)=U(B_{1},B_{2},B_{3};-\omega,-\eta;z), 
\end{eqnarray}
where, on the right-hand side of the first relation, we have 
\begin{eqnarray}
&&B_{1}^{'}=\omega B_{1},  \ 
B_{2}^{'}=2+2i\eta,  \ 
B_{3}^{'}=B_{3}-\left(\frac{B_{2}}{2}+i\eta\right)
\left(\frac{B_{2}}{2}-i\eta-1\right),
\vspace{3mm}\nonumber \\
&&\omega^{'}=1, \ 
i\eta^{'}=\frac{B_{2}}{2}-1, \ \vartheta=\frac{iB_{1}}{2z}. 
\end{eqnarray}
In the third rule it is assumed that we must change the sign of ($\eta,
\omega$) only where these quantities appear explicitly in $S(z)$,
keeping the expressions for the other parameters  unchanged, even if
they depend on $\eta$ and $\omega$. Moreover, note that the
transformation $r_{1}$ generally gives a solution having a region of
convergence different from that of the initial solution, since it
involves an inversion of the independent variable. For brevity, we use
only $r_{1}$ and $r_{2}$, although $r_{3}$ must be employed to furnish 
the full group of solutions.
 
An integral relation can also be used to generate a new solution from a
previously known one. It provides a pair of solutions which has {\it
essentially} the same series coefficients (that is, coefficients which
differ at most by a constant not depending on $n$) but put some
restrictions on the parameters and arguments of the solutions. The
transformation rules applied to that pair afford new ones, having again
the same coefficients and solutions connected to each other by integral
relations.

We denote by $U_{i}^{\infty}$ the series solutions which converge for
$|z|>0$, and by $U_{i}^{0}$ those which converge for $|z|<\infty$, the
subscript $i$ indicating the pair to which the solution belongs. If
there is a phase parameter $\nu$, it appears as a subscript as well,
let us say, $U_{i\nu}^{\infty}$ and $U_{i\nu}^{0}$. For a specified pair,
$U_{i}^{0}$ will result from $U_{i}^{\infty}$ via an integral
transformation. Solutions with a phase parameter are given by two-sided
series in which the summation index $n$ runs from $-\infty$ to
$\infty$, whereas solutions without that parameter are given by
one-sided series ($n\geq 0$). Under certain conditions, the latter
become finite-series solutions which are called quasi-polynomial
solutions, Heun polynomials or quasi-algebraic solutions.

In the next section we find the kernels to the integral
relations used in sections 3 and 4.  In section 3, we use integral 
relations and transformations rules to derive four pairs
of solutions from an asymptotic expansion in the vicinity of
$z=\infty$.  In each pair, one solution is given as an ascending or
descending power series of $z$ and the other as a series of irregular
confluent hypergeometric functions which, for terminating series, may
be written in terms of generalized Laguerre polynomials.

In section 4 we determine the integral relations between expansions in
Coulomb wave functions. We use only expansions in series of irregular
hypergeometric functions, and find that, in each pair without a phase
parameter, one solution may again be expressed as a generalized
Laguerre polynomial. In addition, we verify that finite-series solutions
occur under the same conditions valid for the corresponding solutions
in section 3.

In section 5 we obtain normal forms for the DCHE --- in which there is
no first derivative term --- and examine the two differential equations
which are particular cases of both the DCHE and the GSWE, namely,
\letra
\begin{eqnarray}\label{10}
&&\frac{d^2W_{1}}{du^2}+\left[\theta_{0}+\theta_{1} \cosh(\kappa u)+
\theta_{2}\cosh(2\kappa u)\right]W_{1}=0, \vspace{3mm}\\
&&\frac{d^2W_{2}}{du^2}+\left[\overline{\theta}_{0}+
\overline{\theta}_{1} \sinh(\kappa u)+
\overline{\theta}_{2}\cosh(2\kappa u)\right]W_{2}=0,
\end{eqnarray}
where the first equation represents the Whittaker-Hill equation (WHE)
or the modified WHE depending on whether $\kappa u$ is pure imaginary
or real, respectively. We also discuss the solutions to  the
Schr\"{o}dinger equation for some potentials which give rise to DCHEs,
and specially to these particular cases. Section 6 is devoted to final
comments while, in the Appendix, integrals used to establish integral
relations are written in terms of irregular confluent hypergeometric
functions rather than in terms of Whittaker functions.
%
%
%\newpage
%
%	 SECTION II: INTEGRAL TRANSFORMATION
%
\section*{2. Kernels for integral relations}
Several kernels are possible for integral transformations of the DCHE
\cite{schmidt,slavyanov}, but we only regard those that will be useful
in the subsequent sections. We follow a procedure similar to the one
employed by Schmidt and Wolf \cite{schmidt}, adapted to the form we
have chosen for the DCHE. Thus, if $U(z)$ is a known solution of
equation (\ref{dche}), we seek a new solution $\hat{U}(z)$ given by the
integral relation
\antiletra          
\begin{eqnarray}
\label{integral}
\hat{U}(z)=\int_{t_{1}}^{t_{2}}K(z,t)U(t)dt, 
\end{eqnarray}
where the kernel $K(z,t)$ is determined from \cite{ince}
\letra   
\begin{eqnarray}
\label{kernel}
L_{z}\{K(z,t)\}=\overline{L}_{t}\{K(z,t)\},
\end{eqnarray}
being the operator $L_{z}$ and its adjoint $\overline{L}_{z}$ given by
\begin{eqnarray}
\begin{array}{l}           
\label{kernel2}
L_{z}:=z^2\frac{\partial^{2}}{\partial z^{2}}+\left[B_{1}+B_{2}z\right]
\frac{\partial}{\partial z}+ \left(\omega^{2}z^2-2\omega\eta z\right),
\vspace{5mm}\\
\overline{L}_{t}:= 
t^2\frac{\partial^{2}}{\partial t^{2}}+\left[-B_{1}
+(4-B_{2})t\right]\frac{\partial}{\partial t}+
\left(\omega^{2}t^2-2\omega\eta t+2-B_{2}\right).
\end{array}
\end{eqnarray}
In terms of $L_{z}$, equation (\ref{dche}) reads
\antiletra
\begin{eqnarray}
\label{dche2}
[L_{z}+B_{3}]U(z)=0,
\end{eqnarray}
where $L_{z}$ is now understood as an ordinary differential operator.
In equation (\ref{integral}) we have chosen the contour of integration
as the line joining $t_{1}$ and $t_{2}$, but we assume that these
endpoints depend on $z$ and, consequently, we have to use the formula
\begin{eqnarray}
\label{derivada}
\frac{d}{dz}\int_{t_{1}}^{t_{2}}F(z,t)dt=
\int_{t_{1}}^{t_{2}}\frac{\partial F(z,t)}{\partial z}dt+
F(z,t_{2})\frac{dt_{2}}{dz}
-F(z,t_{1})\frac{dt_{1}}{dz}
\end{eqnarray}
in order to derive the conditions under which $\hat{U}(z)$ is solution
of the DCHE. Hence, applying $L_{z}$ to integral (\ref{integral})
and using equations (\ref{kernel}) and (\ref{derivada}), we find
\letra
\begin{eqnarray}
\label{integral2}
L_{z}\{\hat{U}(z)\}=
\int_{t_{1}}^{t_{2}}\overline{L}_{t}\{K(z,t)\}U(t)dt +Q(z,t_{2})-
Q(z,t_{1}),
\end{eqnarray}
where ($i=1,2$)
\begin{eqnarray}
\label{Q}
Q(z,t_{i}):= 
\left[z^2\frac{d^2t_{i}}{dz^2}+\left(B_{1}+
B_{2}z\right)\frac{d t_{i}}{d z}\right]U(t_{i})K(z,t_{i})+\hspace{5cm}
\nonumber\\
z^2U(t_{i})\left[\frac{\partial K(z,t_{i})}{\partial 
t_{i}}\left(\frac{dt_{i}}{dz}\right)^2+2\frac{\partial K(z,t_{i})}
{\partial z}\frac{d t_{i}}{d z}
\right]+z^2\left(\frac{dt_{i}}{dz}\right)^2\frac{d U(t_{i})}{d t_{i}}K(z,t_{i}).
\end{eqnarray}
The notation $\overline{L}_{t}\{K(z,t)\}U(t)$ in equation
(\ref{integral2}) means that the operator $\overline{L}_{t}$ acts
uniquely on the object inside braces. Next, we integrate equation
(\ref{integral2}) by parts or, equivalently, by using the identity
\antiletra
\begin{eqnarray*}
\label{lagrange}
\overline{L}_{t}\{K(z,t)\}U(t)-K(z,t)L_{t}\{U(t)\}=
\frac{\partial}{\partial t}P(z,t),
\end{eqnarray*}
where $P(z,t)$ is given by
\begin{eqnarray}
\label{P}
P(z,t):= 
t^2\left[U(t)\frac{\partial K(z,t)}{\partial t}
-K(z,t)\frac{\partial U(t)}{\partial t}\right]+
\left[\left(2-B_{2}\right)t -B_{1}\right]U(t)K(z,t).
\label{concomitant}
\end{eqnarray}
Then we find
\begin{eqnarray*}
L_{z}\{\hat{U}(z)\}=
\int_{t_{1}}^{t_{2}}\left[K(z,t)L_{t}\{U(t)\}+
\frac{\partial{P(z,t)}}{\partial{t}}\right]dt+Q(z,t_{2})-Q(z,t_{1}), 
\end{eqnarray*}
or, using equations (\ref{dche2}) and (\ref{integral}), 
\begin{eqnarray}
\label{condition1}
(L_{z}+B_{3})\hat{U}(z)=P(z,t_{2})+Q(z,t_{2})-P(z,t_{1})-Q(z,t_{1}).
\end{eqnarray}
Therefore, when $U(z)$ is a solution of the DCHE, $\hat{U}(z)$
will also be a solution if the integral (\ref{integral}) exist and the
right hand side of equation (\ref{condition1}) vanishes. In fact, we
require that the `integrated terms' vanish when $t\rightarrow t_{i}$,
that is, 
\letra
\begin{eqnarray}
\label{integrability}
Q(z,t_{i})+P(z,t_{i})=0,
\end{eqnarray}
where, due to equations (\ref{Q}) and (\ref{P}),
\begin{eqnarray}
\label{integrability2}
&&P(z,t_{i})+Q(z,t_{i})=\nonumber\\
&&\left[z^2\frac{d^2t_{i}}{dz^2}+
\left(B_{1}+B_{2}z\right)\frac{dt_{i}}
{dz}+\left(2-B_{2}\right)t_{i}-B_{1}\right]K(z,t_{i})U(t_{i})+
\left[z^2\left(\frac{dt_{i}}{dz}\right)^2-(t_{i})^2\right]
\times\nonumber \\
&& K(z,t_{i})\frac{dU(t_{i})}{dt_{i}}+
\left\{2z^2\frac{\partial K(z,t_{i})}{\partial z}\frac{dt_{i}}{dz}
+\left[z^2\left(\frac{dt_{i}}{dz}\right)^2+(t_{i})^2
\right]\frac{\partial K(z,t_{i})}{\partial t_{i}}\right\}U(t_{i}).
\end{eqnarray}
Below, we find two kernels for integral relations and show that, for
these, the right-hand side of (\ref{integrability2}) assumes a very
simple form. From these kernels, two others --- and the corresponding
transformed solutions $\hat{U}(z)$ --- may be obtained by the change
$(\eta,\omega)$ $\rightarrow$ $(-\eta,-\omega)$. These kernels are
necessary to set up integral relations between the solutions derived
from the ones considered in this paper by means of rule $r_{3}$.\\

%	
%			FIRST NUCLEUS
%
\noindent
{\it First kernel:} Performing the substitution
\antiletra
\begin{eqnarray}
K(z,t)=e^{i\omega (z+t)+(B_{1}/z)}\ z^{2-B_{2}}H_{1}(z,t)
\end{eqnarray}
in equation (\ref{kernel}) and supposing that $H_{1}$ depends on $z$ and $t$ by the product $zt$, we find
\begin{eqnarray}\begin{array}{l}
\left(\xi-1\right)\frac{dH_{1}}{d\xi}-
\left(\frac{B_{2}}{2}-i\eta-2\right)H_{1}=0,\ \ 
\xi:=-2i\omega zt/B_{1}\Rightarrow
\vspace{3mm}\\
H_{1}(z,t)=(\xi-1)^{(B_{2}/2)-i\eta-2}=
\left(-\frac{2i\omega}{B_{1}}zt-1\right)^{\frac{B_{2}}{2}-i\eta-2}.
\end{array}
\end{eqnarray}
Thus, we have a first kernel, denoted by $K_{1}(z,t)$, namely,
\begin{eqnarray}
K_{1}(z,t)=e^{i\omega (z+t)+(B_{1}/z)}\ z^{2-B_{2}}
\left(-\frac{2i\omega}{B_{1}}zt-1\right)^{\frac{B_{2}}{2}-i\eta-2}
\end{eqnarray}
which yields
\begin{eqnarray*}
\hat{U}_{1}(z)=e^{i\omega z+(B_{1}/z)}\ z^{2-B_{2}}
\int_{t_{1}}^{t_{2}}dt \left[e^{i\omega t}
\left(-\frac{2i\omega}{B_{1}}zt-1\right)^{\frac{B_{2}}{2}-i\eta-2}
U(t)\right].
\end{eqnarray*}
We rewrite this integral in terms of $\xi$ and integrate from
$\xi_{1}$ to $\xi_{2}$, assuming that these new endpoints are 
constants to be specified later. We get
\begin{eqnarray}
\label{firstkernel}
\hat{U}_{1}(z)=e^{i\omega z+(B_{1}/z)}\ 
z^{1-B_{2}}\int_{\xi_{1}}^{\xi_{2}}d\xi 
\left[e^{-B_{1}\xi/(2z)}\ 
\left(\xi-1\right)^{(B_{2}/2)-i\eta-2}
U\left(-\frac{B_{1}\xi}{2i\omega z}\right)\right].
\end{eqnarray}
On the other hand, by inserting 
\letra
\begin{eqnarray}\label{ti1}
t_{i}(z)=-B_{1}\xi_{i}/(2i\omega z)
\end{eqnarray}
and $K(z,t)=K_{1}(z,t)$ into the right-hand side of the expression
(\ref{integrability2}), we find that the first term becomes
\begin{eqnarray*}
\left[\frac{(B_{1})^2\xi_{i}}{2i\omega z^2}+
\frac{B_{1}\xi_{i}(B_{2}-2)}{i\omega z}-B_{1}\right]
K_{1}(z,t_{i})U(t_{i}),
\end{eqnarray*}
the second term vanishes because
$z^2\left(dt_{i}/dz\right)^2-(t_{i})^2=0$, and the last term reduces to

\begin{eqnarray*}
\left[\frac{(B_{1})^2\xi_{i}(\xi_{i}-2)}{2i\omega z^2}-
\frac{B_{1}\xi_{i}(B_{2}-2)}{i\omega z}+B_{1}\xi_{i}\right]
K_{1}(z,t_{i})U(t_{i}).
\end{eqnarray*}
Therefore we can write
\begin{eqnarray}
\label{firstcondition}
&&P_{1}(z,t_{i})+Q_{1}(z,t_{i})=\left[\frac{(B_{1})^2\xi_{i}}{2i\omega z^2}+
B_{1}\right](\xi_{i}-1)K_{1}(z,t_{i})U(t_{i})\nonumber\\
&&=\left[\frac{(B_{1})^2\xi_{i}}{2i\omega z^2}+B_{1}\right]
z^{2-B_{2}}(\xi_{i}-1)^{(B_{2}/2)-i\eta-1}U(t_{i})
\exp{\left[i\omega z+\frac{B_{1}}{z}-\frac{B_{1}\xi_{i}}{2z}\right]}.
\end{eqnarray}
Now we choose $\xi_{1}=1$ and $\xi_{2}=\infty$ and then we can use one
of the integrals given in the Appendix to integrate equation
(\ref{firstkernel}) for the solutions given in sections 3 and 4.\\

% 
%			SECOND NUCLEUS
%
\noindent
{\it Second kernel}: Accomplishing the substitution
\antiletra
\begin{eqnarray}
K(z,t)=e^{i\omega (z+t)-(B_{1}/t)}t^{B_{2}-2}H_{2}(z,t)
\end{eqnarray}
in equation (\ref{kernel})and supposing that $H_{2}(z,t)$ depends on $z$ and $t$ by the product $zt$, we find
\begin{eqnarray}\begin{array}{l}
\left(\zeta-1\right)\frac{dH_{2}}{d\zeta}+
\left(i\eta+\frac{B_{2}}{2}\right)H_{2}=0, \ \ 
\zeta:=2i\omega zt/B_{1} \Rightarrow
\vspace{3mm}\\
H_{2}(z,t)=(\zeta-1)^{-(B_{2}/2)-i\eta}=
\left(\frac{2i\omega }{B_{1}}zt-1\right)^{-\frac{B_{2}}{2}-i\eta}.
\end{array}
\end{eqnarray}
Hence, the second kernel, $K_{2}(z,t)$, is
\begin{eqnarray}
K_{2}(z,t)=e^{i\omega (z+t)-(B_{1}/t)}\ t^{B_{2}-2}
\left(\frac{2i\omega}{B_{1}}zt-1\right)^{-\frac{B_{2}}{2}-i\eta}
\end{eqnarray}
which implies
\begin{eqnarray*} \hat{U}_{2}(z)=e^{i\omega z}\int_{t_{1}}^{t_{2}}dt
\left[e^{i\omega t-(B_{1}/t)}\    t^{B_{2}-2}
\left(\frac{2i\omega}{B_{1}}zt-1\right)^{-\frac{B_{2}}{2}-i\eta}
U(t)\right], \end{eqnarray*}
or, in terms of $\zeta$,
\begin{eqnarray}
\label{secondkernel}
\hat{U}_{2}(z)=e^{i\omega z}z^{1-B_{2}}\int_{\zeta_{1}}^{\zeta_{2}}d\zeta 
\left[e^{B_{1}\zeta/(2z)-2i\omega z/ \zeta}
 \zeta^{B_{2}-2}\left(\zeta-1\right)^{-(B_{2}/2)-i\eta}
U\left(\frac{B_{1}\zeta}{2i\omega z}\right)\right].
\end{eqnarray}
This time we have
\letra
\begin{eqnarray}
t_{i}=B_{1}\zeta_{i}/(2i\omega z)
\end{eqnarray}
in the condition (\ref{integrability}), and equation (\ref{integrability2})
can be rewritten as
\begin{eqnarray}
\label{secondcondition}
&&P_{2}(z,t_{i})+Q_{2}(z,t_{i})=\left[\frac{(B_{1})^2\zeta_{i}}
{2i\omega z^2}-B_{1}\right](\zeta_{i}-1)K_{2}(z,t_{i})U(t_{i})=
\left[\frac{(B_{1})^2\zeta_{i}}{2i\omega z^2}-B_{1}\right]\times
\nonumber\\
&&\left(\frac{B_{1}\zeta_{i}}{2i\omega z}\right)^{B_{2}-2}
(\zeta_{i}-1)^{1-(B_{2}/2)-i\eta}U(t_{i})
\exp{\left[i\omega z+\frac{B_{1}\zeta_{i}}{2z}-
\frac{2i\omega z}{\zeta_{i}}\right]}.
\end{eqnarray}
We choose $\zeta_{1}=1$ and $\zeta_{2}=\infty$ and, then, we can once
more use one of the integrals of the Appendix to integrate equation
(\ref{secondkernel}).

The approach we have used in this section is similar to the one
employed by Schmidt and Wolf \cite{schmidt}, in which we have regarded
$\xi$ or $\zeta$ as integration variables instead of $t$.  This, in
turn, implies that $t_{1}$ and $t_{2}$ in the the integral
(\ref{integral}) are functions of $z$, as far as the integration
endpoints in the variables $\xi$ or $\zeta$ are taken as constants
throughout sections 3-4. As said before, we shall find
$U(t)=U^{\infty}(t)$ and $ \hat{U}(z)=U^{0}(z)$ in the integral
relations. Then, using the first identification together with the
validity conditions for the integrals, it will be easy to check that
the right-hand sides of equations (\ref{firstcondition}) and
(\ref{secondcondition}) vanish for the solutions given in sections 3
and 4.

%
%
%		 	ASYMPTOTIC SOLUTIONS
%
%\newpage
\section*{3. Solutions derived from an asymptotic expansion}	
The simpler quasi-polynomial solutions to the DCHE are obtained from
the asymptotic expansions in the vicinity of the singular points $0$ or
$\infty$ \cite{schmidt}. In this section, we begin with an asymptotic
representation for $z \rightarrow \infty$ and use the integral
relations and transformation formulae to form four pairs of solutions.
In each pair, one solution is given in terms of a series of ascending
or descending powers of $z$, and the other in terms of a series of
irregular confluent hypergeometric functions. We find the conditions
for obtaining Heun polynomials and verify that, in this case, the
hypergeometric functions degenerate to generalized Laguerre
polynomials.

The starting-point solution is given by
\antiletra
\letra
\begin{eqnarray}
U_{1}^{\infty}(z)=e^{i\omega z}z^{-i\eta-
(B_{2}/2)}\sum_{n=0}^{\infty}b_{n}^{(1)}(-2i\omega z)^{-n},  
\end{eqnarray}
where the recurrence relations for the coefficients $b_{n}^{(1)}$
are, in abbreviated notation,
\begin{eqnarray}\label{recurrence2}
\alpha_{0}b_{1}+\beta_{0}b_{0}=0,  
\ \ \alpha_{n}b_{n+1}+\beta_{n}b_{n}+
\gamma_{n}b_{n-1}=0\ (n\geq1), 
\end{eqnarray}
in which
\begin{eqnarray}
&&\alpha_{n}  =  n+1, \nonumber
\vspace{3mm}\\
&&\beta_{n} =  n\left(n+1+2i\eta\right)+i\omega B_{1}+B_{3}+ 
\left(\frac{B_{2}}{2}+i\eta\right)\left(1+i\eta -
\frac{B_{2}}{2}\right), 
\vspace{.3cm} \\
&&\gamma_{n}  =2i\omega B_{1}\left[n+i\eta+({B_{2}}/{2})-1\right].
\nonumber 
\end{eqnarray}
These relations yield a characteristic equation in terms of the 
infinite continued fraction
\antiletra
\begin{eqnarray}\label{convergence}
\beta_{0}=\frac{\alpha_{0}\gamma_{1}}{\beta_{1}-}\ \frac{\alpha_{1}
\gamma_{2}}
{\beta_{2}-}\ \frac{\alpha_{2}\gamma_{3}}{\beta_{3}-}\cdots. 
\end{eqnarray}
To obtain the foregoing recurrence relations, we perform the substitutions
\begin{eqnarray*}
U_{1}^{\infty}(z)= e^{i\omega z}z^{-i\eta-(B_{2}/2)}Y(y),
\ y=(-2i\omega z)^{-1},
\end{eqnarray*}
and find 
\begin{eqnarray*}
&&y^2\frac{d^2Y}{dy^2}+\left[1+(2+2i\eta) y+2i\omega 
B_{1}y^2\right]\frac{dY}{dy}+
\left[C+2i\omega B_{1}\left(i\eta+\frac{B_{2}}{2}\right)y
\right]Y=0,\\
&&C:=i\omega B_{1}+B_{3}+\left(\frac{B_{2}}{2}+i\eta\right)
\left(1+i\eta-\frac{B_{2}}{2}\right).
\end{eqnarray*}
Then, inserting $Y(y)=\sum_{n=0}^{\infty}b_{n}^{(1)}y^n $ into this
equation and proceeding in the usual form, we obtain the previous
relations.

According to the theory of ordinary differential equations in the
complex domain (see \cite{olver}, chapter 7), the solution
$U_{1}^{\infty}(z)$ is unique (one-valued) within the sector
\begin{eqnarray}\label{sector1}
-\frac{3\pi}{2}<\arg{(-2i\omega z)}<\frac{3\pi}{2}.
\end{eqnarray}
However we still have to show that it converges for $|z|>0$. To
accomplish this, we divide the recurrence relations (\ref{recurrence2}) by
$b_{n}^{(1)}$ and take the limit when $n\rightarrow \infty$.  This
gives
\begin{eqnarray}\label{limit}
\lim_{n\rightarrow \infty}\frac{b_{n+1}^{(1)}}{b_{n}^{(1)}}=
-\frac{2i\omega B_{1}}{n},
\ \mbox{or}\ \ 
\lim_{n\rightarrow \infty}\frac{b_{n+1}^{(1)}}{b_{n}^{(1)}}=-n.
\end{eqnarray}
These limits may as well be derived by using a Perron-Kreuser theorem 
for difference equations \cite{gautschi}. In order to satisfy 
characteristic equation (\ref{convergence}), we have to choose 
the first limit (minimal solution). Thence,
\begin{eqnarray*}
\lim_{n\rightarrow \infty}\frac{b_{n+1}^{(1)}
(-2i\omega z)^{-n-1}}{b_{n}^{(1)}(-2i\omega z)^{-n}}=\frac{B_{1}}{nz}
\end{eqnarray*}
and, therefore, $U_{1}^{\infty}(z)$ converges for any $|z|>0$.

On the other hand, to get a solution $U_{1}^{0}(z)$ convergent in the
neighborhood of $z=0$ we insert $U_{1}^{\infty}(t)$ into the integral
relation (\ref{firstkernel}). This gives
\begin{eqnarray*}
U_{1}^{0}(z):=\hat{U}_{1}(z)\propto 
e^{i\omega z+(B_{1}/z)}z^{1+i\eta-(B_{2}/2)}
\sum_{n=0}^{\infty}b_{n}^{(1)}
\left(\frac{z}{B_{1}}\right)^{n}I_{n}^{(1)}(z),   
\end{eqnarray*}
being
\begin{eqnarray*}
&&I_{n}^{(1)}(z):= \int_{1}^{\infty}d\xi 
\left[e^{-B_{1}\xi/z} 
\left(\xi-1\right)^{(B_{2}/2)-i\eta-2}\xi^{-n-i\eta-(B_{2}/2)}\right]
\stackrel{A1}{=}
\vspace{3mm}\\
&&\Gamma\left(\frac{B_{2}}{2}-i\eta-1\right)e^{-B_{1}/z}
\left(\frac{B_{1}}{z}\right)^{1+n+2i\eta}
U\left(n+i\eta+\frac{B_{2}}{2},n+2+2i\eta,\frac{B_{1}}{z}\right)
\end{eqnarray*}
where we have used integral (\ref{A1}) which is valid if 
\begin{eqnarray}\label{conditionu1}
\Re\left[(B_{2}/2)-i\eta-1\right]>0, \ \   \Re\left({B_{1}}/{z}\right)>0.
\end{eqnarray}
Thus, the sought solution is given, apart from a multiplicative factor, by
\begin{eqnarray}
\label{uo1}
U_{1}^{0}(z)=e^{i\omega z}z^{-i\eta-
(B_{2}/2)}\sum_{n=0}^{\infty}b_{n}^{(1)}
U\left(n+
i\eta+\frac{B_{2}}{2},n+2+2i\eta,\frac{B_{1}}{z}\right),
\end{eqnarray}
where $U(a,b,y)$ denotes the irregular confluent hypergeometric
function \cite{abramowitz}.  Note moreover that, by inserting
$U_{1}^{\infty}$ into equation (\ref{firstcondition}), we have

\begin{eqnarray*}
&&P_{1}(z,t_{i})+Q_{1}(z,t_{i})\propto
\left[\frac{(B_{1})^2\xi_{i}}{2i\omega z^2}+B_{1}\right]
z^{2+i\eta-(B_{2}/2)}\xi_{i}^{-i\eta-(B_{2}/2)}(\xi_{i}-1)^{(B_{2}/2)-i\eta-1}
\hspace{1cm}\\
&&\times \exp{\left[i\omega z-\frac{B_{1}}{z}(\xi_{i}-1)\right]}
\sum_{n=0}^{\infty}b_{n}^{(1)}
\left(\frac{B_{1}\xi_{i}}{z}\right)^{-n}.
\end{eqnarray*}
The series on the right-hand side converges at $\xi_{1}=1$ and at
$\xi_{2}=\infty$. Then, the condition (\ref{conditionu1}) assures that,
for $\xi_{1}=1$, the second member goes to zero since
$(\xi_{1}-1)^{(B_{2}/2)-i\eta-1}\rightarrow 0$; for $\xi_{2}=\infty$,
the second member also vanishes because
$\exp{\left[-B_{1}(\xi_{2}-1)/z\right]}\rightarrow 0$.  Arguments
similar to these may be repeated for the other pairs of solutions
written below.

To obtain the behaviour of $U_{1}^{0}(z)$ when $z\rightarrow 0$, 
we use the relation \cite{erdelyi1} 
\begin{eqnarray}\label{asymptotic}
U(a,b,y)\sim y^{-a}[1+O(|y|^{-1}],\ -3\pi/2<\arg{y}<3\pi/2, \ \ (|y|
\rightarrow \infty),
\end{eqnarray}
thereof we find
\begin{eqnarray}
\lim_{z\rightarrow 0}U_{1}^{0}(z)\sim 1+O\left(\frac{z}{B_{1}}\right),
\ \mbox{within the sector}\ -\frac{3\pi}{2}<\arg{\left(\frac{B_{1}}{z}
\right)}<\frac{3\pi}{2}.
\end{eqnarray}
However, to show that the series in $U_{1}^{0}(z)$ converges for
$|z|<\infty$, we must consider the behaviour of $U(a,b,z)$ when
$b\rightarrow \infty$, while $b-a$ and $z$ remain bounded
\cite{erdelyi1}. In this manner we get
\begin{eqnarray*}
\lim_{n\rightarrow \infty}\frac{f_{n+1}}{f_{n}}=\frac{z}{B_{1}},
\ \ f_{n}:=U\left(n+
i\eta+\frac{B_{2}}{2},n+2+2i\eta,\frac{B_{1}}{z}\right).
\end{eqnarray*}
Then, combining this expression with the first limit given in (\ref{limit}), 
we have
\begin{eqnarray}
\lim_{n\rightarrow \infty}\frac{b_{n+1}^{(1)}f_{n+1}}{b_{n}^{(1)}f_{n}}=
-\frac{2i\omega z}{n}.
\end{eqnarray}
Therefore, $U_{1}^{0}(z)$ converges in any finite region of the complex
plane. In case of finite series, the ratio test becomes meaningless
and the convergence must be decided by inspection.

Starting from the first pair of solutions, we generate three others by
the transformation rules $r_{1}$ and $r_{2}$ and, in each pair, the
solutions are also connected by an integral transformation.  Below, we
collect up the four pairs of solutions $(U_{i}^{\infty},U_{i}^{0})$
($i=1,2,3,4$), the validity conditions for the integral
transformations, the asymptotic behaviour of each solution, and the
sufficient condition to obtain quasi-polynomial solutions. For the
latter solutions, the the functions $U(a,b,z)$ degenerate to an
generalized Laguerre polynomials because the parameter $a$ becomes a
negative integer $-l$, and thus we have \cite{abramowitz}
\begin{eqnarray}
\label{laguerre}
U(-l,1+\alpha,y)=(-1)^{l}l!L_{l}^{\alpha}(y).
\end{eqnarray}
The condition for quasi-polynomial solutions results from the fact that
a series with three-term recurrence relations such as (\ref{recurrence2})
becomes a finite series with $0\leq n\leq N-1$ if $\gamma_{n}=0$
for some $n=N=$positive integer \cite{arscott}.\\

%
%
%
%
%			PRIMEIRO PAR
%
\noindent
{\it First pair} : $(U_{1}^{\infty},U_{1}^{0})$.
\letra
\begin{eqnarray}\label{terminate1}
\begin{array}{l}
U_{1}^{\infty}(z)=e^{i\omega z}z^{-i\eta-
(B_{2}/2)}\sum_{n=0}^{\infty}b_{n}^{(1)}(-2i\omega z)^{-n},   
\vspace{3mm}\\
U_{1}^{0}(z)=e^{i\omega z}z^{-i\eta-
(B_{2}/2)}\sum_{n=0}^{\infty}b_{n}^{(1)}
U\left(n+
i\eta+\frac{B_{2}}{2},n+2+2i\eta,\frac{B_{1}}{z}\right).
\end{array}
\end{eqnarray}
\begin{eqnarray}
\label{terminate1b}
&&\alpha_{n}^{(1)}  =  n+1, \nonumber
\vspace{3mm}\\
&&\beta_{n}^{(1)}  =  n\left(n+1+2i\eta\right)+i\omega B_{1}+B_{3}+ 
\left(\frac{B_{2}}{2}+i\eta\right)\left(1+i\eta -
\frac{B_{2}}{2}\right), 
\vspace{.3cm} \\
&&\gamma_{n}^{(1)}  =2i\omega B_{1}\left(n+i\eta+\frac{B_{2}}{2}-1\right).
\nonumber\\
&&\mbox{Integral relation (\ref{firstkernel})}: 
\Re{\left(\frac{B_{2}}{2}-i\eta-1\right)}>0 \ \mbox{and}\ 
\Re{\left(\frac{B_{1}}{z}\right)}>0. \vspace{3mm}\\ 
&&\mbox{Finite series: }
(B_{2}/2)+i\eta=1-N\ \Rightarrow\ 0\leq n \leq N-1.\hspace{2cm}
\end{eqnarray}
To find the integral relation we have used equation (\ref{A1}).
The asymptotic  behaviours are
\antiletra\letra
\begin{eqnarray}\label{asymptotic1}
&&\lim_{z\rightarrow  \infty}U_{1}^{\infty}(z)\sim e^{i\omega z}z^{-i\eta-
(B_{2}/2)}, \ \ \  -\frac{3\pi}{2}<\arg{(-2i\omega z)}<\frac{3\pi}{2};\\
&&\lim_{z\rightarrow 0}U_{1}^{0}(z)\sim 1,
\hspace{3cm}-\frac{3\pi}{2}<\arg{\left(\frac{B_{1}}{z}\right)}<\frac{3\pi}{2}.
\end{eqnarray}\\
%
%
%			SECOND PAIR
%
%
\noindent
{\it Second pair} : $U_{2}^{\infty}\ \propto \ r_{2}U_{1}^{\infty}$ and 
$U_{2}^{0}\ \propto \ r_{2}U_{1}^{0}$.
\antiletra
\letra
\begin{eqnarray}\label{terminate2}\begin{array}{l}
U_{2}^{\infty}(z)  =e^{i\omega z+(B_{1}/z)}z^{-i\eta-
(B_{2}/2)}\sum_{n=0}^{\infty}b_{n}^{(2)}(-2i\omega z)^{-n},   
\vspace{3mm}\\
U_{2}^{0}(z) =e^{i\omega z+(B_{1}/z)}z^{-i\eta-
(B_{2}/2)}\sum_{n=0}^{\infty}
b_{n}^{(2)}U\left(n+2+i\eta-\frac{B_{2}}{2},n+2+2i\eta,
-\frac{B_{1}}{z}\right).\hspace{1cm}
\end{array}\end{eqnarray}
\begin{eqnarray}
\label{terminate2b}
&&\alpha_{n}^{(2)}  =  n+1, \nonumber
\vspace{3mm}\\
&&\beta_{n}^{(2)}  = n\left(n+1+2i\eta\right)-i\omega B_{1}+B_{3}+ 
\left(\frac{B_{2}}{2}+i\eta\right)\left(1+i\eta -
\frac{B_{2}}{2}\right),
\vspace{3mm}\\
&&\gamma_{n}^{(2)}  =-2i\omega B_{1}\left(n+1+i\eta-\frac{B_{2}}{2})\right).
\nonumber \vspace{3mm}\\
\label{terminate2c}
&&\mbox{Integral relation (\ref{secondkernel}): }
\Re{\left(\frac{B_{2}}{2}+i\eta-1\right)}<0 \ \mbox{and}\ 
\Re{\left(\frac{B_{1}}{z}\right)}<0. \vspace{3mm}\\ 
&&\mbox{Finite series, if }
(B_{2}/2)-i\eta=1+N\ \Rightarrow\ 0\leq n \leq N-1.\hspace{2cm}
\end{eqnarray}
The integral relation is found by using equation (\ref{A1}).  The
asymptotic expressions are 
\antiletra\letra
\begin{eqnarray}\label{asymptotic2}
&&\lim_{z\rightarrow  \infty}U_{2}^{\infty}(z)\sim e^{i\omega z}z^{-i\eta-
(B_{2}/2)}, \ \ \  -\frac{3\pi}{2}<\arg{(-2i\omega z)}<\frac{3\pi}{2};\\
&&\lim_{z\rightarrow 0}U_{2}^{0}(z)\sim e^{B_{1}/z}z^{2-B_{2}},
\hspace{1.4cm}-\frac{3\pi}{2}
<\arg{\left(-\frac{B_{1}}{z}\right)}<\frac{3\pi}{2}.
\end{eqnarray}\\
%
%
%
%				THIRD PAIR
%
\noindent
{\it Third pair} : $U_{3}^{\infty}\ \propto \ r_{1}U_{1}^{0}$ and 
$U_{3}^{0}\ \propto \ r_{1}U_{1}^{\infty}$.
\antiletra
\letra
\begin{eqnarray}\label{terminate3}\begin{array}{l}
U_{3}^{\infty}(z)=e^{i\omega z}
\sum_{n=0}^{\infty}b_{n}^{(3)}
U\left(n+i\eta+\frac{B_{2}}{2},n+B_{2},-2i\omega z\right),   
\vspace{3mm}\\
U_{3}^{0}(z)=e^{i\omega z}\sum_{n=0}^{\infty}b_{n}^{(3)}
\left(\frac{z}{B_{1}}\right)^{n}.
\end{array}\end{eqnarray}
\begin{eqnarray}
\label{terminate3b}
&&\alpha_{n}^{(3)}  =  n+1, \nonumber
\vspace{3mm}\\
&&\beta_{n}^{(3)}  =  n\left(n+B_{2}-1\right)+i\omega B_{1}+B_{3}, 
\vspace{.3cm} \\
&&\gamma_{n}^{(3)}  =2i\omega B_{1}\left(n+i\eta+\frac{B_{2}}{2}-1\right).
\nonumber \vspace{3mm}\\
&&\mbox{Integral relation (\ref{firstkernel}): }
\Re{\left(\frac{B_{2}}{2}-i\eta-1\right)}>0 \ \mbox{and}\ 
\Re{\left(\frac{B_{1}}{z}\right)}>0. \vspace{3mm}\\ 
&&\mbox{Finite series, if }
(B_{2}/2)+i\eta=1-N\ \Rightarrow\ 0\leq n \leq N-1.\hspace{2cm}
\end{eqnarray}
To find the integral relation we use the equation (\ref{A2}). The 
asymptotic behaviours are the same as in the first pair, that is,
\antiletra\letra
\begin{eqnarray}\label{asymptotic3}
&&\lim_{z\rightarrow  \infty}U_{3}^{\infty}(z)\sim e^{i\omega z}z^{-i\eta-
(B_{2}/2)}, \ \ \  -\frac{3\pi}{2}<\arg{(-2i\omega z)}<\frac{3\pi}{2};\\
&&\lim_{z\rightarrow 0}U_{3}^{0}(z)\sim 1,
\hspace{3cm}-\frac{3\pi}{2}<\arg{\left(\frac{B_{1}}{z}\right)}<\frac{3\pi}{2}.
\end{eqnarray}\\
%
%
%			FOURTH PAIR
%
\noindent
{\it Fourth pair} : $U_{4}^{\infty}\ \propto\ r_{2}U_{3}^{\infty}$ and 
$U_{4}^{0}\ \propto \ 
r_{2}U_{3}^{0}$.
\antiletra
\letra
\begin{eqnarray}\label{terminate4}\begin{array}{l}
U_{4}^{\infty}(z) =e^{i\omega z+(B_{1}/z)}
z^{2-B_{2}}\sum_{n=0}^{\infty}
b_{n}^{(4)}U\left(n+2+i\eta-\frac{B_{2}}{2},n+4-B_{2},
-2i\omega z\right),
\vspace{3mm}\\
U_{4}^{0}(z)  =e^{i\omega z+(B_{1}/z)}z^{2-B_{2}}\sum_{n=0}^{\infty}b_{n}^{(4)}
\left(-\frac{z}{B_{1}}\right)^{n}.   
\end{array}\end{eqnarray}
\begin{eqnarray}\label{terminate4b}
&&\alpha_{n}^{(4)}  =  n+1, \nonumber
\vspace{3mm}\\
&&\beta_{n}^{(4)}  =
 n\left(n+3-B_{2}\right)+2-i\omega B_{1}-B_{2} +B_{3},  
\vspace{.3cm} \\
&&\gamma_{n}^{(4)}  =-2i\omega B_{1}\left(n+1+i\eta-\frac{B_{2}}{2}
\right).
\nonumber\vspace{3mm}\\ 
&&\mbox{Integral relation (\ref{secondkernel}): }
\Re{\left(\frac{B_{2}}{2}+i\eta-1\right)}<0 \ \mbox{and}\ 
\Re{\left(\frac{B_{1}}{z}\right)}<0.\vspace{3mm} \\ 
&&\mbox{Finite series, if }
(B_{2}/2)-i\eta=1+N\ \Rightarrow\ 0\leq n \leq N-1.\hspace{2cm}
\end{eqnarray}
To find the integral transformation, we must use relation (\ref{A2}) again.
The asymptotic behaviours are the same as in the second pair, namely,
\antiletra\letra
\begin{eqnarray}\label{asymptotic4}
&&\lim_{z\rightarrow  \infty}U_{4}^{\infty}(z)\sim e^{i\omega z}z^{-i\eta-
(B_{2}/2)}, \ \ \  -\frac{3\pi}{2}<\arg{(-2i\omega z)}<\frac{3\pi}{2},\\
&&\lim_{z\rightarrow 0}U_{4}^{0}(z)\sim e^{B_{1}/z}z^{2-B_{2}},
\hspace{1.4cm}-\frac{3\pi}{2}
<\arg{\left(-\frac{B_{1}}{z}\right)}<\frac{3\pi}{2}.
\end{eqnarray}

For $z\rightarrow 0$, we have found the two asymptotic
behaviours we could expect from the theory of differential equations.
However, for $z\rightarrow \infty$, we have only one of the expected 
expressions. This occurs because we have regarded only one half of
the solutions. In effect, if we apply rule $r_{3}$ to the 
preceding solutions, we get four new solutions $U_{i}^{\infty}(z)$
($i=5,\cdots,8$) for which
\begin{eqnarray*}
\lim_{z\rightarrow  \infty}U_{i}^{\infty}(z)\sim e^{-i\omega z}z^{i\eta-
(B_{2}/2)}, \ \ \  -\frac{3\pi}{2}<\arg{(2i\omega z)}<\frac{3\pi}{2}.
\end{eqnarray*}

Note as well that $U_{3}^{0}(z)$ and $U_{4}^{0}(z)$ can be reexpressed
in terms of confluent hypergeometric functions. Indeed, if we use
integral (A2) to derive these solutions from $U_{3}^{\infty}(z)$ and
$U_{4}^{\infty}(z)$, we find
\begin{eqnarray*}
&&U_{3}^{0}(z)\propto e^{i\omega z}z^{-i\eta-(B_{2}/2)}\sum_{n=0}^{\infty}
b_{n}^{(3)}U\left(n+i\eta+\frac{B_{2}}{2},n+1+i\eta+\frac{B_{2}}{2},
\frac{B_{1}}{z}\right),\\
&&U_{4}^{0}(z)\propto e^{i\omega z+(B_{1}/z)}z^{-i\eta-(B_{2}/2)}
\sum_{n=0}^{\infty}
b_{n}^{(4)}U\left(n+2+i\eta-\frac{B_{2}}{2},n+3+i\eta-\frac{B_{2}}{2},
-\frac{B_{1}}{z}\right),
\end{eqnarray*}
which are consistent with the previous expressions due to the
transformation \cite{abramowitz}
\antiletra
\begin{eqnarray}\label{kummer}
U(a,b,y)=z^{1-b}U(1+a-b,2-b,y)
\end{eqnarray}
followed by (\ref{laguerre}) with $l=0$. From the above expression for 
$U_{3}^{0}(z)$, we find
\begin{eqnarray*}
&&U_{1}^{\infty}(z)=r_{1}U_{3}^{0}(z)\propto e^{i\omega z}\sum_{n=0}^{\infty}
b_{n}^{(1)}U\left(n+i\eta+\frac{B_{2}}{2},n+1+i\eta+\frac{B_{2}}{2},
-2i\omega z\right) \ \Rightarrow\\
&&U_{2}^{\infty}(z)=r_{2}U_{1}^{\infty}(z)\propto e^{i\omega z+(B_{1}/z)}
z^{2-B_{2}}
\sum_{n=0}^{\infty}b_{n}^{(2)}\times\\
&&\hspace{4.8cm}U\left(n+2+i\eta-\frac{B_{2}}{2},
n+3+i\eta-\frac{B_{2}}{2},
-2i\omega z\right),
\end{eqnarray*}
which are alternative forms for $U_{1}^{\infty}(z)$ and
$U_{2}^{\infty}(z)$.  It is useful to write the two solutions in each
pair as series of hypergeometric functions because we can use equation
(\ref{asymptotic}) to deduce the asymptotic behaviours of both
solutions as well as the respective sectors inside which they are
one-valued.

%
%\newpage
\section*{4. Solutions in series of Coulomb wave functions }
%
%
%                   COULOMB COM PARAMETRO
%  
In this section we establish integral relations for expansions in
series of Coulomb wave functions (already found in section 4 of
\cite{eu}) and provide some additional properties for these solutions.
We consider only expansions in series of irregular confluent
hypergeometric functions and, thus, we use the notation
$(U_{i}^{\infty}, U_{i}^{0})$ where we have used $(\widetilde{U}_{i},
U_{i})$ in \cite{eu}. In effect, the solutions in terms of
$U(a,b,z)$ afford the expected behaviour for the solutions when
$z\rightarrow\infty$ and $1/z\rightarrow\infty$.  Moreover, in section
4.2 we note that, in each pair of solutions without a phase parameter,
one solution may be expressed in series of generalized Laguerre
polynomials, and we also find that the conditions for quasi-polynomial
solutions are the same as in the corresponding solutions of section 3.
Therefore, to discard the expansions in regular confluent
hypergeometric functions does not imply that we are setting 
aside finite-series solutions.\\

\noindent
{\it 4.1. Solutions with a phase parameter}\\

\noindent
The first pair below is equivalent to the solutions found by Leaver
\cite{leaver1}. $U_{1\nu}^{0}(z)$ can be derived from
$U_{1\nu}^{\infty}(z)$ by the rule $r_{1}$ and also by an integral
transformation. The second pair results from the first one by means of
the rule $r_{2}$; its solutions are connected to one another by an
integral relation but not by the rule $r_{1}$.\\

%
%			FIRST PAIR \NU
%
\noindent
{\it First pair} :
\letra
\begin{eqnarray}\begin{array}{l}
U_{1\nu}^{\infty}(z) =e^{i\omega z}z^{\nu+1-(B_{2}/2)}
\sum_{n=-\infty}^{\infty}b_{n}(-2i\omega z)^{n}U(n+\nu+1+
i\eta,2n+2\nu+2,-2i\omega z),
\vspace{0.3cm}\\
U_{1\nu}^{0}(z)  =e^{i\omega z}z^{-\nu-(B_{2}/2)}\sum_{n=-\infty}^{\infty}b_{n}
\left(\frac{B_{1}}{z}\right)^{n}U\left(n+\nu+\frac{B_{2}}{2},2n+2\nu+2,
\frac{B_{1}}{z}\right),
\end{array}\end{eqnarray}
with the following recurrence relations for the coefficients 
$b_{n}$
\begin{eqnarray}
\alpha_{n}b_{n+1}+\beta_{n}b_{n}+\gamma_{n}b_{n-1}=0, \ \ 
%\alpha_{n}^{'}b_{n+1}^{'}+\beta_{n}^{'}b_{n}^{'}+\gamma_{n}^{'}b_{n-1}^{'}=0,
%\label{nu8}
\end{eqnarray}
where
\begin{eqnarray}
&&\begin{array}{l}
\alpha_{n}  =  \frac{i\omega B_{1}
[n+\nu+2-(B_{2}/2)][n+\nu+1-i\eta]}
{2[n+\nu+1][n+\nu+(3/2)]},
\vspace{.3cm} \\
\beta_{n}  = B_{3}+\left(n+\nu+1-\frac{B_{2}}{2}\right)\left(n+\nu+\frac{B_{2}}{2}\right)
+ \frac{\eta \omega B_{1}[(B_{2}/2)-1]}
{(n+\nu)(n+\nu+1)},
\vspace{0.3cm} \\
\gamma_{n}=\frac{i\omega B_{1} [n+\nu+(B_{2}/2)-1][n+\nu+
i\eta]}{2[n+\nu][n+\nu-(1/2)]}. 
\end{array}\\
&&\mbox{Integral relation (\ref{firstkernel}): }
\Re{[(B_{2}/2)-i\eta-1)}>0, \ \ \Re{(B_{1}/z)}>0.
\end{eqnarray} 
The phase parameter $\nu$ may be determined from a characteristic 
equation given as a sum of two infinite continued fractions, namely,
\antiletra
\begin{eqnarray}
\label{nu12}
\beta_{0}=\frac{\alpha_{-1}\gamma_{0}}{\beta_{-1}-} \frac{\alpha_{-2}
\gamma_{-1}}{\beta_{-2}-}\frac{\alpha_{-3}\gamma_{-2}}
{\beta_{-3}-}\cdots+\frac{\alpha_{0}\gamma_{1}}{\beta_{1}-} 
\frac{\alpha_{1}\gamma_{2}}
{\beta_{2}-}\frac{\alpha_{2}\gamma_{3}}{\beta_{3}-}\cdots .
\end{eqnarray}
To show that these solutions are connected to each other by the integral
relation mentioned above, we introduce $U(t) =U_{1\nu}^{\infty}(t)$ into
the integral (\ref{firstkernel}).  Then we find
\begin{eqnarray*}
\hat{U}_{1}(z)\propto e^{i\omega z+(B_{1}/z)}z^{-\nu-(B_{2}/2)}
\sum_{n=-\infty}^{\infty}b_{n}
\left(\frac{B_{1}}{z}\right)^{n}I_{n}^{(1)}(z)\ \propto \ 
U_{1\nu}^{0}(z)  
\end{eqnarray*}
since ($\xi:=-2i\omega zt/B_{1}$)
\begin{eqnarray*}
I_{n}^{(1)}=\int_{1}^{\infty} 
e^{-B_{1}\xi/z}\ 
\left(\xi-1\right)^{(B_{2}/2)-i\eta-2}\xi^{n+\nu+1-(B_{2}/2)}
U\left(n+\nu+1+
i\eta,2n+2\nu+2,\frac{B_{1}}{z}\xi\right)d\xi
\vspace{3mm}\\
\stackrel{A3}{=}\Gamma\left(\frac{B_{2}}{2}-i\eta-1\right)
e^{-B_{1}/z}U\left(n+\nu+\frac{B_{2}}{2},2n+2\nu+2, \frac{B_{1}}{z}
\right),\hspace{4.3cm}
\end{eqnarray*}
under the conditions written in (48d). Furthermore, by inserting
$U_{1\nu}^{\infty}(z)$ into condition (\ref{firstcondition}), we find
($t_{i}=-B_{1}\xi_{i}/(2i\omega z)$)
\begin{eqnarray*}
&&P_{1}(z,t_{i})+Q_{1}(z,t_{i})\propto\left[\frac{(B_{1})^2\xi_{i}}
{2i\omega z^2}+B_{1}\right]
z^{1-\nu-(B_{2}/2)}\xi_{i}^{1+\nu-(B_{2}/2)}(\xi_{i}-1)^{(B_{2}/2)-i\eta-1}
\hspace{1cm}\\
&&\times \exp{\left[i\omega z-\frac{B_{1}}{z}(\xi_{i}-1)\right]}
\sum_{n=0}^{\infty}b_{n}^{(1)}
\left(\frac{B_{1}\xi_{i}}{z}\right)^{n}U\left(n+\nu+1+
i\eta,2n+2\nu+2,\frac{B_{1}}{z}\xi_{i}\right).
\end{eqnarray*}
As in section 3, the right-hand side of this expression vanishes
for $\xi_{1}=1$ and $\xi_{2}=\infty$ and, therefore, the condition
(\ref{integrability}) is satisfied.\\

%
%
%			SECOND PAIR \NU
%
\noindent
{\it Second pair} :
\letra
\begin{eqnarray}
\begin{array}{l}
U_{2\nu}^{\infty}(z) =f(z)z^{\nu+1-
(B_{2}/2)}\sum_{n=-\infty}^{\infty}b_{n}^{'}
(-2i\omega z)^{n}U(n+\nu+1+
i\eta,2n+2\nu+2,-2i\omega z),\vspace{3mm}\\
U_{2\nu}^{0}(z)  =f(z)z^{-\nu-
(B_{2}/2)}\sum_{n=-\infty}^{\infty}b_{n}^{'}
\left(-\frac{B_{1}}{z}\right)^{n}U\left(n+\nu+2-
\frac{B_{2}}{2},2n+2\nu+2,
-\frac{B_{1}}{z}\right),
\vspace{3mm}\\
\hspace{6cm}f(z):=e^{i\omega z+(B_{1}/z)},
\end{array}\end{eqnarray}
where the recurrence relations for $b_{n}^{'}$ are
\begin{eqnarray}
\label{nu18}
\alpha_{n}^{'}b_{n+1}^{'}+\beta_{n}^{'}b_{n}^{'}+
\gamma_{n}^{'}b_{n-1}^{'}=0,
\end{eqnarray}
with
\begin{eqnarray}
&&\begin{array}{l}
\alpha_{n}^{'}  =  \frac{i\omega B_{1}
[n+\nu+(B_{2}/2)][n+\nu+1-i\eta]}
{2[n+\nu+1][n+\nu+(3/2)]}, \vspace{0.3cm} \\
\beta_{n}^{'} = -B_{3}-\left(n+\nu+1-\frac{B_{2}}{2}\right)
\left(n+\nu+\frac{B_{2}}{2}\right)
- \frac{\eta \omega B_{1}[(B_{2}/2)-1]}
{(n+\nu)(n+\nu+1)},
\vspace{0.3cm} \\
\gamma_{n}^{'}= \frac{i\omega B_{1}[n+\nu+1-(B_{2}/2)][n+\nu+
i\eta]}{2[n+\nu][n+\nu-(1/2)]}.
\end{array} \\
&&\mbox{Integral relation (\ref{secondkernel}): }\Re{[(B_{2}/2)+
i\eta-1)}<0, \ \ \Re{(B_{1}/z)}<0.
\end{eqnarray} 
The characteristic equation is analogous to (\ref{nu12}). 
Again, it is simple to find the integral relation stated above. We 
insert $U_{2\nu}^{\infty}(t)$ into the right-hand side of 
relation (\ref{secondkernel}) and get 
\begin{eqnarray*}
\hat{U}_{2}(z)\propto e^{i\omega z}z^{-\nu-(B_{2}/2)}
\sum_{n=-\infty}^{\infty}b_{n}^{'}
\left(-\frac{B_{1}}{z}\right)^{n}I_{n}^{(2)}(z)\propto \ U_{2\nu}^{0}(z),
\end{eqnarray*}
since ($\zeta:=2i\omega zt/B_{1}$)
\begin{eqnarray*}
\begin{array}{l}
I_{n}^{(2)}=\int_{1}^{\infty} e^{B_{1}\zeta/z}\
 \left(\zeta-1\right)^{-(B_{2}/2)-i\eta}
\zeta^{n+\nu-1+(B_{2}/2)}
U\left(n+\nu+1+i\eta,2n+2\nu+2,-\frac{B_{1}}{z}\zeta\right)d\zeta
\vspace{3mm}\\
\hspace{1cm}\stackrel{A3}{=}\Gamma\left(1-i\eta-\frac{B_{2}}{2}
\right)
e^{B_{1}/z}U\left(n+\nu+2-\frac{B_{2}}{2},2n+2\nu+2,
-\frac{B_{1}}{z}\right).
\end{array}
\end{eqnarray*}
On the other hand, by inserting $U_{2\nu}^{\infty}$ into the equation
(\ref{secondcondition}) we find that the condition
(\ref{integrability}) is satisfied.  Moreover, we note that the
validity conditions for the integrals do not involve the phase
parameter and, therefore, these integral relations are also valid for
the truncated solutions.\\

%
%
%		COULOMB WITHOUT PHASE PARAMETER
%
%
\noindent
{\it 4.2. Solutions without phase parameter}\\
\noindent
These come from the truncation of the solutions with phase parameter
($n\geq 0$) but, contrary to the solutions given in section 3, now
there are three possible forms for the recurrence relations and for
the corresponding characteristic equations (see appendix of
\cite{eu}).  For completeness, we write out these relations. In the
first one we have $\alpha_{-1}=0$ and, in the other cases, $\alpha_{-1}$
may be different from zero.
\antiletra
\begin{eqnarray}
&&\left.
\begin{array}{l}
\alpha_{0}b_{1}+\beta_{0}b_{0}=0,  
\vspace{.2cm} \\ 
\alpha_{n}b_{n+1}+\beta_{n}b_{n}+
\gamma_{n}b_{n-1}=0\ (n\geq1),
\end{array}\right\} \Rightarrow
\beta_{0}=\frac{\alpha_{0}\gamma_{1}}{\beta_{1}-}\ \frac{\alpha_{1}
\gamma_{2}}
{\beta_{2}-}\ \frac{\alpha_{2}\gamma_{3}}{\beta_{3}-}\cdots. \hspace{28mm}
\label{r1a}
\end{eqnarray}
\begin{eqnarray}
\left.
\begin{array}{l}
\alpha_{0}b_{1}+\beta_{0}b_{0}=0, 
\vspace{.2cm} \\
\alpha_{1}b_{2}+\beta_{1}b_{1}+\left[
\alpha_{-1}+\gamma_{1}\right]b_{0}=0, 
\vspace{.2cm} \\ 
\alpha_{n}b_{n+1}+\beta_{n}b_{n}+\gamma_{n}
b_{n-1}=0\ (n\geq2),
\end{array}\right\}\Rightarrow
\beta_{0}=\frac{\alpha_{0}\left[\alpha_{-1}+\gamma_{1}
\right]}
{\beta_{1}-} 
\ \frac{\alpha_{1}\gamma_{2}}{\beta_{2}-}\ \frac{\alpha_{2}\gamma_{3}}
{\beta_{3}-}\cdots .
\label{r2a}
\end{eqnarray}
\begin{eqnarray}
\left.
\begin{array}{l}
\alpha_{0}b_{1}+\left[\beta_{0}+\alpha_{-1}
\right]b_{0}=0, 
\vspace{.2cm} \\
\alpha_{n}b_{n+1}+\beta_{n}b_{n}
+\gamma_{n}b_{n-1}=0\ (n\geq1),
\end{array}\right\}\Rightarrow 
\beta_{0}+\alpha_{-1}=\frac{\alpha_{0}\gamma_{1}}
{\beta_{1}-}\ \frac{\alpha_{1}\gamma_{2}}{\beta_{2}-}
\ \frac{\alpha_{2}\gamma_{3}}
{\beta_{3}-}\cdots .\hspace{3mm}
\label{r3a}
\end{eqnarray}

In each one of the the four pairs of truncated solutions, one solution
can be expressed in terms of generalized Laguerre polynomials by the
Kummer transformation (\ref{kummer}) followed by equation
(\ref{laguerre}).  Such solutions are: $U_{1}^{0}(z)$, $U_{2}^{0}(z)$,
$U_{3}^{\infty}(z)$ and $U_{4}^{\infty}(z)$. In the case of
finite-series solutions, the other solutions may as well be written as
series of generalized Laguerre polynomials, since we have $a=$zero or a
negative integer in $U(a,b,y)$.

In addition to have changed the notations, we have also reordered the
solutions of \cite{eu} so that the pairs are obtained  by using 
rules $r_{2}$ and $r_{1}$ in the same sequence as in section 3.
Thereupon we find that the integral relations, the conditions for
having terminating series as well as the asymptotic behaviour of each
solution are the same as those in the corresponding pairs of section 3. \\

%
%                       PRIMEIRO PAR
%
\noindent
{\it First pair} : 
$\nu=i\eta\ \mbox{in}\ (U_{1\nu}^{\infty},U_{1\nu}^{0})$
\letra
\begin{eqnarray}\begin{array}{l}
U_{1}^{\infty}(z) =e^{i\omega z}z^{1+i\eta-(B_{2}/2)}
\sum_{n=0}^{\infty}b_{n}^{(1)}
(-2i\omega z)^{n}U(n+1+2i\eta,2n+2+2i\eta,-2i\omega z),
\vspace{.3cm}\\
U_{1}^{0}(z)=e^{i\omega z}z^{-i\eta-(B_{2}/2)}\sum_{n=0 }^{\infty}
b_{n}^{(1)}\left(\frac{B_{1}}{z}\right)^{n}U\left(n+
i\eta+\frac{B_{2}}{2},2n+2+2i\eta ,
\frac{B_{1}}{z}\right).
\end{array}\end{eqnarray}
\begin{eqnarray}
&&\alpha_{n}^{(1)}  = \frac{i\omega B_{1}
[n+1]\left[n+2+i\eta-(B_{2}/2)\right]}
{2\left[n+1+i\eta\right]\left[n+i\eta+(3/2)\right]},\nonumber
\vspace{.3cm} \\
&&\beta_{n}^{(1)}  = B_{3}+\left(n+1+i\eta-\frac{B_{2}}{2}\right)
\left(n+i\eta+\frac{B_{2}}{2}\right)
+\frac{\eta \omega B_{1}\left[(B_{2}/2)-1\right]}
{\left(n+i\eta\right)\left(n+1+i\eta\right)},
\vspace{.3cm} \\
&&\gamma_{n}^{(1)}= \frac{i\omega B_{1}\left[n+2i\eta\right]\left[n+
(B_{2}/2)+i\eta-1\right]}{2\left[n+i\eta\right]
\left[n+i\eta-(1/2)\right]}.\nonumber 
\end{eqnarray} 
\begin{eqnarray}
&&\mbox{Integral relation (\ref{firstkernel}): }
\Re{[(B_{2}/2)-i\eta-1]}>0 \ \mbox{and}\ \Re{(B_{1}/z)}>0. \\ 
&&\mbox{Recurrence relations}:
\left\{
\begin{array}{ll}
\mbox{Eq. (\ref{r1a}),}& \mbox{if} \ i\eta\neq0,-1/2; \\
\mbox{Eq. (\ref{r2a}),}&\mbox{if} \  i\eta=-1/2; \\
\mbox{Eq. (\ref{r3a}), }&\mbox{if} \  i\eta=0.
\end{array}
\right.\\
&&\mbox{Finite series, if }
(B_{2}/2)+i\eta=1-N\ \Rightarrow\ 0\leq n \leq N-1.
\end{eqnarray}
The asymptotic behavior is given by equations (\ref{asymptotic1}-b).\\

%
%
%
%
%                       SEGUNDO PAR 
%
%
%

\noindent
{\it Second pair} : 
$\nu=i\eta$ in ($U_{2\nu}^{\infty},U_{2\nu}^{0}$) or 
$U_{2}^{\infty}\ \propto \ r_{2}U_{1}^{\infty}$ and $U_{2}^{0}\ 
\propto \ r_{2}U_{1}^{0}$.
\antiletra
\letra
\begin{eqnarray}\begin{array}{l}
U_{2}^{\infty}(z) =f(z)z^{1+i\eta-(B_{2}/2)}
\sum_{n=0}^{\infty}b_{n}^{(2)}
(-2i\omega z)^{n}U(n+1+2i\eta,2n+2+2i\eta,-2i\omega z),
\vspace{3mm}\\
U_{2}^{0}(z) =f(z)z^{-i\eta-(B_{2}/2)}
\sum_{n=0}^{\infty}b_{n}^{(2)}
\left(-\frac{B_{1}}{z}\right)^{n}U\left(n+2+i\eta-
\frac{B_{2}}{2},2n+2+2i\eta,
-\frac{B_{1}}{z}\right),\vspace{3mm}\\
\hspace{6cm}f(z):=e^{i\omega z+(B_{1}/z)}.
\end{array}\end{eqnarray}
\begin{eqnarray}
&&\alpha_{n}^{(2)}  = -\frac{i\omega B_{1}
[n+1]\left[n+(B_{2}/2)+i\eta\right]}
{2\left[n+1+i\eta\right]\left[n+i\eta+(3/2)\right]} ,\nonumber
\vspace{.3cm} \\
&&\beta_{n}^{(2)}  = B_{3}+\left(n+1+i\eta-\frac{B_{2}}{2}\right)
\left(n+i\eta+\frac{B_{2}}{2}\right)
+\frac{\eta \omega B_{1}\left[(B_{2}/2)-1\right]}
{\left(n+i\eta\right)\left(n+1+i\eta\right)},
\vspace{.3cm} \\
&&\gamma_{n}^{(2)}= -\frac{i\omega B_{1}\left[n+2i\eta\right]\left[n-
(B_{2}/2)+i\eta+1\right]}{2\left[n+i\eta\right]
\left[n+i\eta-(1/2)\right]} .\nonumber
\end{eqnarray} 
\begin{eqnarray}
&&\mbox{Integral relation (\ref{secondkernel}): }
\Re{[(B_{2}/2)+i\eta-1]}<0 \ \mbox{and}\ \Re{(B_{1}/z)}<0. \\
%\end{eqnarray}
%
&&\mbox{Recurrence relations}:
\left\{
\begin{array}{ll}
\mbox{Eq. (\ref{r1a}),}& \mbox{if} \ i\eta\neq0,-1/2; \\
\mbox{Eq. (\ref{r2a}),}&\mbox{if} \  i\eta=-1/2; \\
\mbox{Eq. (\ref{r3a}), }&\mbox{if} \  i\eta=0.
\end{array}
\right.\\
&&\mbox{Finite series, if }
(B_{2}/2)-i\eta=N+1\ \Rightarrow\ 0\leq n \leq N-1.
\end{eqnarray}
The asymptotic behavior is given by equations (\ref{asymptotic2}-b).\\

%%
%                       TERCEIRO PAR 
%
%
\noindent
{\it Third pair} : $ \nu=B_{2}/2-1$ in 
$(U_{1\nu}^{\infty},U_{1\nu}^{0}$) or
$U_{3}^{\infty}\ \propto \ r_{1}U_{1}^{0}$ and 
$U_{3}^{0}\ \propto \ r_{1}U_{1}^{\infty}$.
\antiletra
\letra
\begin{eqnarray}\begin{array}{l}
U_{3}^{\infty}(z) =e^{i\omega z}\sum_{n=0}^{\infty}b_{n}^{(3)}
(-2i\omega z)^{n}U\left(n+\frac{B_{2}}{2}+i\eta,2n+B_{2},
-2i\omega z\right),
\vspace{.3cm}\\ 
U_{3}^{0}(z) =e^{i\omega z}z^{1-B_{2}}\sum_{n=0}^{\infty}b_{n}^{(3)}
\left(\frac{B_{1}}{z}\right)^{n}U\left(n+B_{2}-1,2n+B_{2},
\frac{B_{1}}{z}\right).
\end{array}\end{eqnarray}
\begin{eqnarray}
&&\alpha_{n}^{(3)}  =  \frac{i\omega B_{1}
[n+1]\left[n+(B_{2}/2)-i\eta\right]}
{2\left[n+(B_{2}/2)\right]\left[n+(B_{2}/2)+
(1/2)\right]},\nonumber
\vspace{.3cm} \\
&&\beta_{n}^{(3)}  = B_{3}+n(n+B_{2}-1)+
\frac{\eta \omega B_{1}\left[(B_{2}/2)-1\right]}
{\left[n+(B_{2}/2)-1\right]\left[n+(B_{2}/2)\right]},
\vspace{.3cm} \\
&&\gamma_{n}^{(3)}= \frac{i\omega B_{1}\left[n+B_{2}-2\right]\left[n+
(B_{2}/2)-1+i\eta\right]}{2\left[n+(B_{2}/2)-1\right]
\left[n+(B_{2}/2)-(3/2)\right]}.\nonumber
\end{eqnarray} 
\begin{eqnarray}
&&\mbox{Integral relation (\ref{firstkernel}): }
\Re{[(B_{2}/2)-i\eta-1]}>0 \ \mbox{and}\ \Re{(B_{1}/z)}>0. \\ 
&&\mbox{Recurrence relations}:
\left\{
\begin{array}{ll}
\mbox{Eq. (\ref{r1a}),}& \mbox{if} \ B_{2}\neq 1,2; \\
\mbox{Eq. (\ref{r2a}),}&\mbox{if} \  B_{2}=1; \\
\mbox{Eq. (\ref{r3a}), }&\mbox{if} \  B_{2}=2.
\end{array}
\right.\\
&&\mbox{Finite series, if }
(B_{2}/2)+i\eta=1-N\ \Rightarrow\ 0\leq n \leq N-1.\hspace{2cm}
\end{eqnarray}
The asymptotic behavior is given by equations (\ref{asymptotic3}-b).\\

%
%                       QUARTO PAR 
%
\noindent
{\it Fourth pair} : $\nu=1-B_{2}/2$ in
($U_{2\nu}^{\infty}$,$U_{2\nu}^{0}$) or 
$U_{4}^{\infty}\ \propto\ r_{2}U_{3}^{\infty}$ and $U_{4}^{0}\ \propto \ 
r_{2}U_{3}^{0}$. 
\antiletra
\letra
\begin{eqnarray}\begin{array}{l}
U_{4}^{\infty}(z) =f(z)z^{2-B_{2}}
\sum_{n=0}^{\infty}b_{n}^{(4)}
(-2i\omega z)^{n}U\left(n+2-\frac{B_{2}}{2}+i\eta,2n+4-B_{2},
-2i\omega z\right),
\vspace{.3cm}\\
U_{4}^{0}(z) =f(z)z^{-1}
\sum_{n=0}^{\infty}b_{n}^{(4)}
\left(-\frac{B_{1}}{z}\right)^{n}U\left(n+3-B_{2},2n+4-B_{2},
-\frac{B_{1}}{z}\right),
\vspace{3mm}\\ 
\hspace{6cm}f(z):=e^{i\omega z+(B_{1}/z)}.
\end{array}\end{eqnarray}
\begin{eqnarray}
&&\alpha_{n}^{(4)}  =  \frac{i\omega B_{1}
[n+1]\left[n+2-(B_{2}/2)-i\eta\right]}
{2\left[n+2-(B_{2}/2)\right]\left[n+(5/2)-
(B_{2}/2)\right]},\nonumber
\vspace{.3cm} \\
&&\beta_{n}^{(4)}  =- B_{3}-(n+1)(n+2-B_{2})
-\frac{\eta \omega B_{1}\left[(B_{2}/2)-1\right]}
{\left[n+1-(B_{2}/2)\right]\left[n+2-(B_{2}/2)\right]},
\vspace{.3cm} \\
&&\gamma_{n}^{(4)}= \frac{i\omega B_{1}\left[n+2-B_{2}\right]\left[n+
1-(B_{2}/2)+i\eta\right]}{2\left[n+1-(B_{2}/2)\right]
\left[n+({1}/{2})-(B_{2}/2)\right]}.\nonumber
\end{eqnarray} 
\begin{eqnarray}
&&\mbox{Integral relation (\ref{secondkernel}): }
\Re{[(B_{2}/2)+i\eta-1]}<0 \ \mbox{and}\ \Re{(B_{1}/z)}<0. \\ 
&&\mbox{Recurrence relations}:
\left\{
\begin{array}{ll}
\mbox{Eq. (\ref{r1a}),}& \mbox{if} \ B_{2}\neq 2,3; \\
\mbox{Eq. (\ref{r2a}),}&\mbox{if} \  B_{2}=3; \\
\mbox{Eq. (\ref{r3a}), }&\mbox{if} \  B_{2}=2.
\end{array}
\right.\\
&&\mbox{Finite series, if }
(B_{2}/2)-i\eta=N+1\ \Rightarrow\ 0\leq n \leq N-1.
\end{eqnarray}
The asymptotic behavior is given by equations (\ref{asymptotic4}-b).

We remark that, in addition to the three possible forms for the
recurrence relations, the coefficients of the latter are fractional.
Thence, these relations are not well defined when some denominator
vanishes. Thus, if $i\eta=$ negative integer or half-integer$<-1/2$,
the coefficients of the first and second pairs are not well defined,
but we can form well-defined expressions by using the rule $r_{3}$
which changes $(\eta, \omega)$ by $(-\eta, -\omega)$.  Similarly, if
for some value of $B_{2}$ a denominator vanishes for the the third
pair, we must consider the solutions of the fourth pair, and
vice-versa.

%
%			PARTICULAR CASES
%
%\newpage
%			%
\section*{5. DCHE and GSWE: special cases and examples}
In this section we examine the two differential equations (\ref{10}-b)
which share the property of being particular cases of both the DCHE and
the GSWE.  In such equations, namely,
\antiletra
\begin{eqnarray}\label{special1}
&&\frac{d^2W_{1}}{du^2}+\left[\theta_{0}+\theta_{1} \cosh(\kappa u)+
\theta_{2}\cosh(2\kappa u)\right]W_{1}=0, \\ %
&&\frac{d^2W_{2}}{du^2}+\left[\overline{\theta}_{0}+
\overline{\theta}_{1} \sinh(\kappa u)+
\overline{\theta}_{2}\cosh(2\kappa u)\right]W_{2}=0, \end{eqnarray} 
$\kappa$ is a given constant such that $\kappa u$ is real or pure
imaginary and the $\theta_{i}\ (\overline{\theta}_{i})$ are constants.
Thus there are only three parameters in each equation.  If $\kappa u$
is pure imaginary, the first is the Whittaker-Hill equation (WHE) and
if $\kappa u$ is real, the modified WHE \cite{arscott}. In fact,
Decarreau, Maroni and Robert \cite{decarreaux2} have already found
that the WHE has that property, whereas we  have found that these two
equations are special cases of the GSWE \cite{eu}.  Now we find some
normal forms for the DCHE, from one of these we get the particular
equations written above and show that they also come from a GSWE with
$B_{2}=1$ and $B_{1}=-z_{0}/2$.  Finally, we discuss solutions for
problems obeying DCHE and GSWE, intending to decide on the best
interpretation for each of the special equations (\ref{special1}-59).
\\

%
%				NORMAL FORMS
%
\noindent
{\it 5.1. Normal forms for the DCHE}\\

\noindent
Several normal forms for general Heun's equation and its confluent 
cases (except the triconfluent equation) were established by Lemieux 
and Bose \cite{lemieux}. Below, we give some forms suitable for DCHE
when it is written as in equation (\ref{dche}). 

By performing the substitution
\letra
\begin{eqnarray}
U(z)=z^{-B_{2}/2}e^{B_{1}/(2z)}F(z)
\end{eqnarray}
in equation (\ref{dche}), we find for $f(z)$ an algebraic normal form of the
DCHE , namely,
\begin{eqnarray}\label{normal1}
\frac{d^2F}{dz^2}+\left[\omega^{2}-\frac{2\eta \omega}{z}+\frac{1}{z^2}
\left(B_{3}-\frac{B_{2}^{2}}{4}+\frac{B_{2}}{2}\right)+\frac{B_{1}}{z^3}
\left(1-\frac{B_{2}}{2}\right)
-\frac{B_{1}^2}{4z^4}\right]F=0, 
\end{eqnarray}
where, as before, $B_{1}\neq 0,\ \omega\neq 0$. The further transformations
\antiletra\letra
\begin{eqnarray}\label{normal}
z=e^{\lambda u},\ \ F(z)=e^{\lambda u/2}W(u)\ \Rightarrow
\ W(u)=z^{(B_{2}-1)/2}e^{-B_{1}/(2z)}U(z), 
\end{eqnarray}
where $\lambda$ is a constant at our disposal, bring the equation 
to the hyperbolic normal form 
\begin{eqnarray}\label{normal2}
&&\frac{d^2W}{du^2}+\lambda^2I(u)W=0,\\
&&I(u):=-\left[B_{1}\left(1-\frac{B_{2}}{2}\right)+2\eta \omega\right]
\sinh(\lambda u)+\left[\omega^2+\frac{B_{1}^2}{4}\right]\sinh(2\lambda u)+
\nonumber\\
&&
\left[B_{1}\left(1-\frac{B_{2}}{2}\right)-2\eta \omega\right]
\cosh(\lambda u)+\left[\omega^2-\frac{B_{1}^2}{4}\right]
\cosh(2\lambda u)+B_{3}-\frac{1}{4}\left(1-B_{2}\right)^2,
\hspace{1cm}
\end{eqnarray}
to be used soon. The last form we shall need is obtained by the 
transformations
\antiletra
\letra
\begin{eqnarray}
z= \rho^2, \ \ U(z)=\rho^{(1-2B_{2})/2}e^{B_{1}/(2\rho^2)}G(\rho) 
\Rightarrow G(\rho)=z^{(2B_{2}-1)/4} e^{-B_{1}/(2z)}U(z)
\end{eqnarray}
in equation (\ref{dche}), and this affords another algebraic normal form
given by
\begin{eqnarray}\label{normal4}
\frac{d^2G}{d\rho^2}+\left[4\omega^{2}\rho^2-8\eta \omega+\frac{4}{\rho^2}
\left(B_{3}-\frac{B_{2}^{2}}{4}+\frac{B_{2}}{2}-\frac{3}{16}\right)+
\frac{4B_{1}}{\rho^4}
\left(1-\frac{B_{2}}{2}\right)
-\frac{B_{1}^2}{\rho^6}\right]G=0. 
\end{eqnarray}\\
%
%			SPECIAL CASES
%
\noindent
{\it 5.2. The common special cases}\\

\noindent
Equation (\ref{normal2}-c) gives the relations among parameters which
lead to equations (\ref{special1}-60) as special cases of the DCHE. For
the WHEs we have
\antiletra
\begin{eqnarray}\label{eqn2}
\begin{array}{l}
\omega^2=-\frac{B_{1}^2}{4},\ 
 2\eta \omega=-B_{1}\left[1-\frac{B_{2}}{2}\right]\Rightarrow
\vspace{.3cm}\\
\frac{d^2W_{1}}{du^2}+\lambda^2\left[B_{3}-\frac{1}{4}
\left(1-B_{2}\right)^2-4\eta \omega \cosh(\lambda u)+
2\omega^{2}\cosh(2\lambda u)\right]W_{1}=0,
\end{array}
\end{eqnarray}
and for the second equation 
\begin{eqnarray}\label{eqn3}
\begin{array}{l}
\omega^2=-\frac{B_{1}^2}{4},\ 
 2\eta \omega=B_{1}\left[1-\frac{B_{2}}{2}\right]\Rightarrow
\vspace{.3cm}\\
\frac{d^2W_{2}}{du^2}+\lambda^2\left[B_{3}-\frac{1}{4}
\left(1-B_{2}\right)^2-4\eta \omega \sinh(\lambda
u)+2\omega^{2}\cosh(2\lambda u)\right]W_{2}=0.
\end{array}
\end{eqnarray}
Now we consider the particular GSWE  
\begin{eqnarray}
&&z(z-z_{0})\frac{d^{2}U}{dz^{2}}+
\left(-\frac{z_{0}}{2}+z\right)\frac{dU}{dz}+ 
\left[B_{3}-2\eta \omega (z-z_{0})+
\omega^{2}z(z-z_{0})\right]U=0,\hspace{1cm}\\
&&\hspace{3cm}[B_{1}=-{z_{0}}/{2}, \ B_{2}=1\ 
\mbox{in equation (4)}]\nonumber
\end{eqnarray}
which has only three constants, since $z_{0}$ may be chosen at will,
excepting zero. Then, the two special cases follow from the last
equation by a change in the independent variable. For the WHEs we have
\begin{eqnarray}\label{var1}
\begin{array}{l}
z=z_{0}\cosh^{2}(\sigma u/2), \ \ U(z)=W_{1}(u)\Rightarrow
\vspace{3mm}\\
\frac{d^2W_{1}}{du^2}+\sigma^2\left[B_{3}+\eta\omega z_{0}
-\frac{1}{8}\omega^2z_{0}^2-
\eta\omega z_{0}\cosh(\sigma u)+
\frac{1}{8}\omega^{2}z_{0}^{2}\cosh(2\sigma u)\right]W_{1}=0,
\end{array} 
\end{eqnarray}
and for the second equation
\begin{eqnarray}\label{var2}
\begin{array}{l}
z=\frac{z_{0}}{2}\left[i\sinh(\sigma u)+1\right], \ \ U(z)=W_{2}(u)
\Rightarrow \vspace{3mm}\\
\frac{d^2W_{2}}{du^2}+\sigma^2\left[B_{3}+\eta \omega z_{0}-
\frac{1}{8}\omega^2 z_{0}^2
-i\eta \omega z_{0} \sinh(\sigma u)-\frac{1}{8}\omega^2 z_{0}^2
\cosh(2\sigma u)\right]W_{2}=0.
\end{array}
\end{eqnarray}
Note that the symbols $B_{i}$, $\eta$ and $\omega$ are denoting 
different objects depending on whether they appear in the DCHE or 
in the GSWE.

In general, solutions for the WHE, when obtained from solutions for the
GSWE, are even or odd with respect to the variable $u$. For example:
(i) we have found pairs of even or odd solutions constituted by one
solution in series of hyperbolic or trigonometric functions ---
Arscott's solutions \cite{arscott,arscott1}--- and another in series of
Coulomb wave functions \cite{eu} , (ii) the Hylleraas and Jaff\'e type
solutions to the GSWE \cite{leaver1} yield only even solutions for the
WHE, but the transformation rule $T_{2}$ given in equation (6b) of
\cite{eu} generates new solutions whose limits are odd. \\

%
%				EXAMPLES
%
\noindent
{\it 5.3. The examples}\\

\noindent
We write the one-dimensional time-independent Schr\"{o}dinger equation 
for a particle with mass $m$ and energy $E$ as
\begin{eqnarray} 
\label{schr}
\frac{d^2\psi(u)}{du^2}+[{\cal E}-V(u)]\psi(u)=0, \ \ u:=a x, \ 
 \ \ {\cal E}:=\frac{2mE}{\hbar^2 a^2},
\end{eqnarray}
where $a$ is a constant, $x$ the spatial Cartesian coordinate, and
$\psi$ must satisfy the regularity conditions 
\begin{eqnarray}
\label{regularity}
\lim_{u\rightarrow \pm \infty }\psi=0.
\end{eqnarray}
For the two following examples, we will see that it is convenient
to interpret the WHE as a GSWE and the second special equation as a DCHE.\\

%
%			FIRST EXAMPLE
%
\noindent
{\it First example}: For the symmetric double-Morse potential of
Zaslavskii and Ulyanov \cite{zaslavskii}
\begin{eqnarray}\label{zas}
V(u)=\frac{B^2}{4}\sinh^2{u} 
-B\left(s+\frac{1}{2}\right)\cosh{u}, \ \ B>0,
\end{eqnarray}
the Schr\"{o}dinger equation is a modified WHE.  If $s$ is a
non-negative integer or a half-integer, the potential (\ref{zas}) is a
quasi-exactly solvable (QES) potential in the sense that one part of
the energy spectrum stems from finite-series solutions
\cite{usheveridze}. If this WHE is considered as a GSWE we find
\cite{eu}:  (i) even and odd quasi-polynomial solutions satisfying the
regularity conditions, and (ii) even and odd infinite-series solutions
which also satisfy the regularity conditions, provided that we match
solutions having the same series coefficients and different radii of
convergence.  However, if the WHE is interpreted as a DCHE, only
finite-series solutions satisfy the regularity conditions, as we will
see next.

For the sake of generality, we will get the solutions for the symmetric
potential (\ref{zas}) as limits of the solutions for the asymmetric
potential \cite{zaslavskii}
\begin{eqnarray}\label{zas1}
V(u)=\frac{B^2}{4}\left(\sinh{u}-\frac{C}{B}\right)^{2} 
-B\left(s+\frac{1}{2}\right)\cosh{u},  (B>0, C\geq 0)
\end{eqnarray}
for which the Schr\"{o}dinger equation becomes the DCHE 
\begin{eqnarray} 
\label{schr2}\frac{d^2\psi(u)}{du^2}+\left[\frac{CB}{2}\sinh u+
B\left(s+\frac{1}{2}\right)\cosh u-
\frac{B^2}{8}\cosh(2u)+{\cal E}-\frac{C^2}{4}+\frac{B^2}{8}\right]\psi(u)=0. 
\end{eqnarray}
Comparing the above equation with equations (\ref{normal}-c), we find 
\letra
\begin{eqnarray}
\lambda=1\ \Rightarrow z=e^u,\ \psi(u)=W(u)=z^{(B_{2}-1)/2}
e^{-B_{1}/(2z)}U(z),
\end{eqnarray}
and the parameters 
\begin{eqnarray}
\label{paramzas}
B_{1}=\frac{B}{2},\  B_{2}=1+C-2s,\ B_{3}={\cal E}+\frac{B^2}
{8}+s^2-sC, \ 
\omega=i\frac{B}{4},\ i\eta=-\frac{C}{2}-\frac{1}{2}-s. 
\end{eqnarray}
To form regular solutions we identify $U(z)$ with the first and the
third pairs of solutions given in section 3.  The first pair yields
%
%				FIRST
%
\antiletra\letra
\begin{eqnarray}\begin{array}{l} 
\psi_{1}^{\infty}(u)=e^{-\frac{B}{2}\cosh{u}
+(\frac{C}{2}+s)u}
\sum_{n=0}^{\infty}
b_{n}^{(1)} \left(
\frac{B}{2}e^{u}\right)^{-n},\vspace{3mm}\\
\psi_{1}^{0}(u)=e^{-\frac{B}{2}\cosh{u}+
\left(\frac{C}{2}+s\right)u}
\sum_{n=0}^{\infty}
b_{n}^{(1)} U\left(n-2s, n+1-C-2s,
 \frac{B}{2}e^{-u}\right),\hspace{1cm}
\end{array}\end{eqnarray}
with the following coefficients in the recurrence relations
(\ref{recurrence2}) for $b_{n}^{(1)}$
\begin{eqnarray}
&&\alpha_{n}^{ (1)}=-(n+1), \nonumber\\
&&\beta_{n}^{ (1)}=-{\cal E}-s(s+C)-n(n-C-2s)= 
\overline{\beta}_{n}^{\ (1)}-{\cal E},
\\				
&&\gamma_{n}^{ (1)}=\frac{B^2}{4}(n-2s-1).  \nonumber
\end{eqnarray}
If $s$ is a
non-negative integer or half-integer (QES potential), we have
$\gamma_{2s+1}=0$ and thus the series are finite with $0\leq
n\leq 2s$. Thence,  the recurrence relations can be written as
\antiletra
\begin{eqnarray}
\label{matrix}
\left(
\begin{array}{cccccccc}
\overline{\beta}_{0} & \alpha_{0} &    0       & \cdots   &     &  &   &     0  \\
\gamma_{1}&\overline{\beta}_{1}   & \alpha_{1} &  0      &    &  &  & \vdots \\  
    0    &\gamma_{2} & \overline{\beta}_{2}    &\alpha_{2}&     &  &   &        \\
 \vdots  &           &            &            &     &  &   &        %\\ 
 %        &    &   &   &       &\gamma_{2s-1}&\overline{\beta}_{2s-1} %&\alpha_{2s-1}
\\         
         &    &   &   &       &      0      & \gamma_{2s}
&\overline{\beta}_{2s}
\end{array}
\right) \left(\begin{array}{l}
b_{0}  \\
 \vdots   \\
        \\
       \\ 
b_{2s-1}\\
b_{2s}\end{array}
\right)=
{\cal E}
\left(\begin{array}{l}
b_{0}  \\
 \vdots   \\
        \\
       \\ 
b_{2s-1}\\
b_{2s}\end{array}\right).
\end{eqnarray}
This system of equations determines $2s+1$ different and real values
for ${\cal E}$ since, for a eigenvalue problem like this, that is, with
a tridiagonal matrix, the following theorem holds (see \cite{arscott},
page 21):  `if $\alpha_{j}$, $\overline{\beta}_{j}$, $\gamma_{j}$ are
real and each product $\alpha_{j} \gamma_{j+1}$ is positive, then the
roots corresponding to the equation (\ref{matrix}) are all real and
different'. In fact, this remains valid even if both $\alpha_{j}$ and
$\gamma_{j}$ are pure imaginaries, since we can take $c_{n}=i^nb_{n}$
and put $\alpha_{j}^{'}=-i\alpha_{j}$, $\beta_{n}^{'}=\beta_{n}$,
$\gamma_{j}^{'}=i\gamma_{j}$ in the recurrence relations for $c_{n}$.
The second condition of the theorem stands for the present case because
\begin{eqnarray*}
\alpha_{j} \gamma_{j+1}=-(j+1)(j-2s)B^2/4>0,
\end{eqnarray*}
where the inequality follows from the fact that $0\leq j \leq 2s-1$ for
the elements of that matrix. Note that for these quasi-polynomial
expansions we can select either the solution $\psi_{1}^{\infty}(u)$ or
$\psi_{3}^{0}(\xi)$, since the hypergeometric functions in the latter
reduce to a generalized Laguerre polynomial. The same is true for the
solutions that result from the third pair of section 3,
%
%				SEcond solution
%
\letra
\begin{eqnarray}\begin{array}{l} 
\psi_{3}^{\infty}(u)=e^{-\frac{B}{2}\cosh{u}
+(\frac{C}{2}-s)u}
\sum_{n=0}^{\infty}
b_{n}^{(3)} U\left(n-2s, n+1+C-2s,
 \frac{B}{2}e^{u}\right),\vspace{3mm}\\
\psi_{3}^{0}(\xi)=e^{-\frac{B}{2}\cosh{u}+
\left(\frac{C}{2}-s\right)u}
\sum_{n=0}^{\infty}
b_{n}^{(3)} \left(
 \frac{B}{2}e^{-u}\right)^{-n},\hspace{1cm}
\end{array}\end{eqnarray}
for which the coefficients of the recurrence relations for $b_{n}^{(3)}$ 
are 
\begin{eqnarray}
&&\alpha_{n}^{ (3)}=-(n+1), \nonumber\\		
&&\beta_{n}^{ (3)}=-{\cal E}-s(s-C)-n(n+C-2s)=
\overline{\beta}_{n}^{\ (3)}-{\cal E}, 
\\				
&&\gamma_{n}^{ (3)}=\frac{B^2}{4}(n-2s-1).  \nonumber
\end{eqnarray}
The above pairs of solutions are related to one another by means of
\antiletra
\begin{eqnarray}
\label{symmetry}
\left(\psi_{1}^{\infty}(C;u),
\psi_{1}^{0}(C;u)\right) \leftrightarrow
\left(\psi_{3}^{0}(-C;-u),\psi_{3}^{\infty}(-C;-u)
\right),
\end{eqnarray}
that is, these pairs remain invariant under the change
$(C,u)\leftrightarrow (-C,-u)$, a property also present in
Schr\"{o}dinger equation (\ref{schr2}). Note that we have obtained in
\cite{eu} one pair of finite-series solutions for equation
(\ref{schr2}) by using the first pair of solutions given in section
4.2; another pair might be obtained from that by using equation
(\ref{symmetry}) or, alternatively, by using the third pair of
solutions given in section 4.2.  However, such solutions have
recurrence relations with fractional coefficients and, for this reason,
they are not well defined for certain integer values of the parameter
$C$, as noted there.

If we suppose that $s$ is not a non-negative integer or half-integer,
we can --- following Leaver \cite{leaver1}, section 8 --- match the
infinite-series solutions of each pair to get regular eigenfunctions 
which converge over the entire range of $u$, since the
$\psi_{1}^{\infty}$ converge when $\exp{u}>0$ and the the
$\psi_{1}^{0}$ converge when $\exp{u}<\infty$. In this case, the energy
spectrum may be determined from the infinite continued fraction
(\ref{convergence}). Note also that the conditions to get
$\psi_{i}^{0}$ from $\psi_{i}^{\infty}$ by an integral transformation
are $\Re{(C)}>0$ and $\Re{(Be^{-u})}>0$ and, therefore, are assured.

On the other hand, by using the second and fourth pairs of solutions
given in section 3 we may form two pairs of infinite-series solutions
(even if the potential is QES) but these are not regular when
$u\rightarrow -\infty$ due to a multiplicative factor
$\exp{[-(B/2)\sinh{u}}]$.

For the symmetric case ($C=0$) the two  pairs degenerate to only one  
\letra
\begin{eqnarray}\begin{array}{l} 
\psi_{1}^{\infty}(u)=e^{-\frac{B}{2}\cosh{u}
+su}
\sum_{n=0}^{\infty}
b_{n}^{(1)} \left(
\frac{B}{2}e^{u}\right)^{-n},\vspace{3mm}\\
\psi_{1}^{0}(u)=e^{-\frac{B}{2}\cosh{u}+su}
\sum_{n=0}^{\infty}
b_{n}^{(1)} U\left(n-2s, n+1-2s,
 \frac{B}{2}e^{-u}\right)
\vspace{3mm}\\
\hspace{1.4cm}=(B/2)^{2s} e^{-\frac{B}{2}\cosh{u}
-su}
\sum_{n=0}^{\infty}
b_{n}^{(1)} \left(
\frac{B}{2}e^{-u}\right)^{-n} \ \ [\mbox{see equation} \ (\ref{kummer})]
\end{array}\end{eqnarray}
and, in the recurrence relations (\ref{matrix}), we have
\begin{eqnarray}
\alpha_{n}^{ (1)}=-(n+1), \ 
\beta_{n}^{(1)}=-{\cal E}-s^2-n(n-2s)= 
\overline{\beta}_{n}^{\ (1)}-{\cal E}, \ 
\gamma_{n}^{(1)}=\frac{B^2}{4}(n-2s-1)  
\end{eqnarray}
For finite-series (QES potential) the two solutions are 
convergent and regular for $u\in(-\infty,\infty)$, but are
neither even nor odd with respect to $u$. However, we can 
form even and odd eigenfunctions by taking the linear combinations 
\begin{eqnarray*}
A_{1}\psi_{1}^{0}(u)+A_{2}\psi_{1}^{\infty}(u)=
e^{-(B/2)\cosh{u}}\sum_{n=0}^{2s}
b_{n}^{(1)} \left[A_{1}\left(\frac{B}{2}\right)^{2s}e^{(n-s)u}+
A_{2}e^{-(n-s)u}\right],
\end{eqnarray*}
and choosing the constants $A_{1}$ and $A_{2}$ such that we have series
of $\cosh{[(n-s)u]}$ and $\sinh{[(n-s)u]}$, respectively. Note the
absence of infinite-series solutions for this QES potential, contrary
to the solutions resulting from the GSWE \cite{eu}.\\

%
%			SECOND EXAMPLE
%
\noindent
{\it Second example}. This example also deals with a QES potential, 
now giving a differential equation of the second type. In contrast with 
the the first example, we find that: (i) if the equation is treated 
as a DCHE,  only infinite-series solutions satisfy the regularity 
conditions, (ii)  if the equation is treated as a GSWE, there is no 
regular solutions.

The potential is
\antiletra
\begin{eqnarray}\label{secondexample}
V(u)=\frac{B^2}{4}\sinh^{2}u-\left(s+\frac{1}{2}\right)B\sinh u,\ \ B>0, \
s=0,\frac{1}{2},1,\frac{3}{2},\cdots
\end{eqnarray}
and thus the Schr\"{o}dinger equation (\ref{schr}) reads
\begin{eqnarray}\label{secondexample2} 
\frac{d^2\psi(u)}{du^2}+\left[{\cal E}+\frac{B^2}{8}+
\left(s+\frac{1}{2}\right)B\sinh u-\frac{B^2}{8}\cosh(2 u)
\right]\psi(u)=0. 
\end{eqnarray}
Comparing this equation with equations (\ref{normal}) and (\ref{eqn3}), 
we get 
\letra
\begin{eqnarray}
\lambda=1\ \Rightarrow z=e^u,\ \psi(u)=W(u)=z^{(B_{2}-1)/2}e^{-B_{1}/(2z)}U(z),
\end{eqnarray}
together with the following expressions for the parameters 
\begin{eqnarray}
\label{paratoy}
B_{1}=-\frac{B}{2}, \ B_{2}=1-2s,\ B_{3}={\cal E}+\frac{B^2}
{8}+s^2, \ 
i\omega=-\frac{B}{4},\ i\eta=-\frac{1}{2}-s. 
\end{eqnarray}
Now, using the second and fourth pairs of solutions given section 3,
we obtain two pair of infinite-series solutions for $\psi(u)$.
The first is
\antiletra\letra
\begin{eqnarray}\begin{array}{l} 
\psi_{2}^{\infty}(u)=e^{-\frac{B}{2}\cosh{u}
+su}
\sum_{n=0}^{\infty}
b_{n}^{(2)} \left(
\frac{B}{2}e^{u}\right)^{-n},\vspace{3mm}\\
\psi_{2}^{0}(u)=e^{-\frac{B}{2}\cosh{u}+su}
\sum_{n=0}^{\infty}
b_{n}^{(2)} U\left(n+1, n+1-2s,
 \frac{B}{2}e^{-u}\right),\hspace{1cm}
\end{array}\end{eqnarray}
with
\begin{eqnarray}
\alpha_{n}^{(1)}=n+1, \ \ 
\beta_{n}^{(1)}={\cal E}+s^2+n(n-2s), \ \ 
\gamma_{n}^{(1)}=-\frac{B^2}{4}
n. 
\end{eqnarray}
in the recurrence relations (\ref{recurrence2}). The second pair 
is given by
\antiletra\letra				
\begin{eqnarray}
\begin{array}{l} 
\psi_{4}^{\infty}(u)=e^{-\frac{B}{2}\cosh{u}
+(1+s)u}
\sum_{n=0}^{\infty}
b_{n}^{(4)} U\left(n+1, n+3+2s,
 \frac{B}{2}e^{u}\right),\vspace{3mm}\\
\psi_{4}^{0}(\xi)=e^{-\frac{B}{2}\cosh{u}+
\left(1+s\right)u}
\sum_{n=0}^{\infty}
b_{n}^{(4)} \left(
 \frac{B}{2}e^{-u}\right)^{-n},\hspace{1cm}
\end{array}\end{eqnarray}
and has
\begin{eqnarray}
\alpha_{n}^{(4)}=n+1, \ \ 		
\beta_{n}^{(4)}={\cal E}+(s+1)^2+n(n+2+2s),\ \  
\gamma_{n}^{(4)}=-\frac{B^2}{4}n. 
\end{eqnarray}
in the recurrence relations (\ref{recurrence2}). In these two pairs of
nonterminating series we have to match the solutions in order to assure
regularity over the entire interval for $u$. Now, the conditions to
derive $\psi_{i}^{0}$ from $\psi_{i}^{\infty}$ by an integral
transformation are $\Re{(2s+1)}>0$ and $\Re{(Be^{-u})}>0$.
Quasi-polynomial solutions result from the first and third pairs of
section 3 but they are not regular due to the presence of the factor
$\exp{[-(B/2)\sinh{u}}]$, as in the previous example.

We have not found any regular solution by describing this problem by a
GSWE. In effect, comparing equations (\ref{secondexample2}) and
(\ref{var2}), we find
\begin{eqnarray*}
\sigma=z_{0}=1, \ z=\frac{i}{2}\sinh{u}+\frac{1}{2}, \
\omega=B, \  i\eta =-s-\frac{1}{2},\ B_{3}={\cal E}+\frac{B^2}{4}
-i\left(s+\frac{1}{2}\right)B.
\end{eqnarray*}
However the solutions for the GSWE have the factor $\exp{(\pm
i\omega z)}=$ $\pm B(-\sinh{u}+i)/2$ which diverges when $u\rightarrow
\infty$ or $u\rightarrow -\infty$ . As an illustration, we use
the pair of solutions given in equation (42) of \cite{eu} and obtain
\begin{eqnarray*}
&&\psi_{1}=
e^{-(B\sinh{u})/2}\sum_{n=0}^{\infty}b_{n}^{(1)}F\left(-n,
n;\frac{1}{2};\frac{1-i\sinh{u}}{2}\right), 
\vspace{3mm}\\ 
&&\widetilde{\psi}_{1} =e^{-(B\sinh{u})/2}\sum_{n=0}^{\infty}
b_{n}^{(1)}(B\sinh{u}-iB)^{n}  
U\left(n-s,2n+1;B\sinh{u}-iB\right),
\end{eqnarray*}
where the recurrence relations have the form given in equations (\ref{r2a}),
with
\begin{eqnarray*}
\alpha_{n}^{ (1)}  =  \frac{iB}{2}\left(n+s+1 \right), \
\beta_{n}^{ (1)}  =-n^2 -{\cal E}-\frac{B^2}{4} , \
\gamma_{n}^{ (1)}  = -\frac{iB}{2}\left(n-s-1 \right). 
\end{eqnarray*} 
Hence it follows that, if $s$ is a non-negative integer, we have finite
series and, if $s$ is a half-integer, we have infinite series, but none
satisfies the regularity conditions.

We note that the potential (\ref{secondexample}) was obtained from
the potential (\ref{zas}) by means of the change $[s+(1/2)]B\cosh u$ 
$\rightarrow$$[s+(1/2)]B\sinh u$, but it can also be derived from the 
potential 
\antiletra
\begin{eqnarray}\label{gonz}
V(u)=V_{1}\sinh^{2} u+V_{2}\sinh u+ 
\frac{V_{3}\sinh u+V_{4}}{\cosh^{2} u}, \ u\in (-\infty, \infty)
\end{eqnarray}
for which the Schr\"{o}dinger equation is a GSWE \cite{lemieux}. A QES 
version of (\ref{gonz}) is given in \cite{gonz} and from that we 
can obtain (\ref{secondexample}) by taking $V_{3}=V_{4}=0$. 

To finalize we mention two other problems which are reducible to a 
DCHE in one or another of the algebraic normal forms. They are related 
with the three-dimensional radial Schr\"{o}dinger equation 
\begin{eqnarray} 
\label{schr3}
\frac{d^2R}{dr^2}+\frac{2}{r}\frac{dR}{dr}+\left[{\cal E}-\frac{l(l+1)}
{r^2}-V(r)\right]R=0, 
 \ \ {\cal E}:=\frac{2mE}{\hbar^2 },
\end{eqnarray}
for a particle with mass $m$ and energy $E$. A normal form for this 
equation is
\begin{eqnarray}\label{radial} 
\frac{d^2H(r)}{dr^2}
+\left[{\cal E}-\frac{l(l+1)}{r^2}-V(r)\right]H(r)=0,\ \ H(r):=rR(r).
\end{eqnarray}
Then, for the inverse fourth-power potential
\begin{eqnarray}
V(r)=V_{1}{r^{-1}}+{V_{2}}{r^{-2}}+{V_{3}}{r^{-3}}+
{V_{4}}{r^{-4}}, 
\end{eqnarray}
equation (\ref{radial}) assumes the normal form (\ref{normal1}) for
the DCHE with $z=r$, whereas for the even-power potential
\begin{eqnarray}
V(r)={V_{1}}{r^{2}}+{V_{2}}{r^{-2}}+{V_{3}}{r^{-4}}+
{V_{4}}{r^{-6}} 
\end{eqnarray}
it assumes the normal form (\ref{normal4}) with $\rho=r$. Solutions
have been proposed for problems like these [20-23], but it would be
intersting to study such problems from the viewpoint of the DCHE since
the transformations rules and integral relations allow us to obtain new
solutions from a known one. We could also check whether the solutions
presented here are useful for that purpose.
%
%
%			CONCLUSIONS
%
\section*{6. Concluding remarks}
We have found integral relations for solutions of the double-confluent
Heun equation and combined them with transformation rules in order to
obtain the group of solutions given in section 3. In section 4, the
integral relations (\ref{firstkernel}) and (\ref{secondkernel}) have
also been used to connect expansions in series of Coulomb wave
functions.  The solutions have been displayed in pairs
($U_{i}^{\infty}(z),U_{i}^{0}(z)$), where $U_{i}^{\infty}(z)$ and
$U_{i}^{0}(z)$ converge for $|z|>0$ and $|z|<\infty$, respectively.
The solutions in each pair have the same series coefficients just
because $U_{i}^{0}(z)$ comes from $U_{i}^{\infty}(z)$ by an integral
relation, whose validity conditions (\ref{integrability}-b) were
satisfied due to the choice of an integration contour that excluded
the point $t=0$, where the solution $U_{i}^{\infty}(t)$ does not
converge.

Comparing each pair of solutions given in section 3 with the
corresponding pair given in section 4.2, we find (i) the same integral
relations between the solutions, (ii) the same asymptotic behaviour for
the solutions, (iii) the same conditions for quasi-polynomial
solutions, and (iv) the possibility of writing one solution as a
generalized Laguerre polynomial.  The last property may be important if
we need to normalize solutions.  Note, however, that in section 3 there
is only one form for the recurrence relations, whereas in section 4.2
there are three possible forms with fractional coefficients which 
are not well defined when a denominator vanishes.

In section 5 we have given normal forms for the DCHE and analysed
equations (\ref{10}-b) that have the common property of being
particular cases of both the DCHE and the GSWE. We have as well looked
for solutions to the Schr\"{o}dinger equation for two quasi-exactly
solvable hyperbolic potentials. For the symmetric Zaslavskii-Ulyanov
potential we have a modified WHE, for which we have established regular
quasi-polynomial and infinite-series solutions by considering this WHE
as a GSWE \cite{eu}, but only quasi-polynomial solutions by treating it
as a DCHE. For the potential (\ref{secondexample}), in the second
example, the Schr\"{o}dinger equation is an equation of the second
special type, for which we have found regular infinite-series solutions
by interpreting the equation as a DCHE; however we have not found any
regular solution by considering it as a GSWE.  For infinite-series
solutions, which have been formed by joining solutions belonging to a
same pair, the energy spectra may be computed from infinite continued
fractions, for instance.

The results of the preceding paragraph suggest that we must interpret
the WHE (10a) as a GSWE, and equation (10b) as a DCHE. However, it
is necessary to consider other problems such as the time
dependence of the Dirac equation in radiation-dominated 
Friedmann-Robertson-Walker spacetimes (see section 2.2.1 of \cite{eu}).
The differential equations for these problems have no free parameters, 
which implies that we have to deal with double-sided series possessing
a phase parameter, as those of section 4.1 for equation (10b)
or the solutions of section 2.2 of \cite{eu} for the WHE.

In section 5 we have also discussed solutions to the Schr\"{o}dinger
equation with an asymmetric Zaslavskii-Ulyanov potential, but now using
the new solutions found in section 3.  The regular solutions are
quasi-polynomial and do not exclude integral values for the parameter
$C$, contrary to the solutions constructed on the basis of the
expansions given in section 4.2 \cite{eu}. Despite this advantage, the
obtainment of regular infinite-series solutions for this problem
remains unsolved.

Finally, we have called attention to some singular radial potentials for
which the Schr\"{o}dinger equation leads to DCHEs, but we have not
tried to solve these equations.  Another related problem concerns the
solutions of the Schr\"{o}dinger equation for the QES potentials
derived recently by Bagchi and Ganguly \cite{bagchi}, since for their
hyperbolic potentials we will find modified WHEs, and for trigonometric
potentials, WHEs. Considering these equations as particular cases of
the GSWE, we may expect to find quasi-polynomial and infinite-series
solutions as in the first example of section 5.

In the Appendix we have rewritten some integrals in a form suitable for
use in section 3 and 4. Thus, excepting a correction to a misprint in a
table of integrals, this appendix does not contain anything original.
%
%                               APENDICE
%
\section*{Appendix. Integrals used in sections 3 and 4}
\protect\label{A}
\setcounter{equation}{0}
\renewcommand{\theequation}{A\arabic{equation}}
The first equation is an integral representation for the irregular
confluent hypergeometric function $U(a,b,z)$ given by 
\cite{abramowitz}
\begin{eqnarray}
\label{A1}
\int_{1}^{\infty}e^{-yt}(t-1)^{\alpha-1}t^{\beta-\alpha-1}dt=
\Gamma(\alpha)e^{-y}U(\alpha,\beta,y), \ \ [\Re{\alpha}>0, \ \ \Re{y}>0].
\end{eqnarray}
The two following integrals are usually given in terms of irregular
Whittaker functions $W_{\kappa,\mu}(z)$. By convenience, we have
reexpressed them in terms of irregular confluent hypergeometric
functions by using the relation \cite{erdelyi1}
\begin{eqnarray*}
W_{\kappa,\mu}(y)=e^{-y/2}y^{\mu+(1/2)}U\left(\frac{1}{2}-\kappa+\mu,
2\mu+1,y\right).
\end{eqnarray*}
The first of these integrals is 
\begin{eqnarray}
\label{A2}
&&\int_{1}^{\infty}
e^{-ay}(y-1)^{\mu-1}U\left(\frac{1}{2}-\kappa-\lambda,
1-2\lambda,ay\right)dy
\vspace{3cmm}\nonumber
\\
&&=\Gamma(\mu)e^{-a}a^{-\mu}
U\left(\frac{1}{2}-\kappa-\lambda,1-2\lambda-\mu,a\right), 
\ \ [\Re{\mu}>0, \ \Re{a}>0],
\end{eqnarray}
which results from \cite{prudnikov}
\begin{eqnarray*}
\int_{1}^{\infty}
e^{-a y/2}(y-1)^{\mu-1}y^{\lambda-\frac{1}{2}}W_{\kappa,\lambda}(a y)
dy
%\vspace{3cmm}\nonumber\\
%
=\Gamma(\mu)e^{-a/2}a^{-\mu/2}
W_{\kappa-\frac{\mu}{2},\ \lambda+\frac{\mu}{2}}(a), 
 \ [\Re{\mu}>0,\ \Re{a}>0].
\end{eqnarray*}
Note a misprint on page 867 of \cite{gradshteyn} where we have
$W_{\kappa-(\mu/2),\ \lambda-(\mu/2)}(a)$ on the right-hand side of the
above integral. The other integral is
\begin{eqnarray}
\label{A3}
&&\int_{1}^{\infty}
e^{-ay}(y-1)^{\mu-1}y^{\kappa+\lambda-\mu-\frac{1}{2}}
U\left(\frac{1}{2}+\lambda-\kappa,2\lambda+1,ay\right)dy
\vspace{3cmm}\nonumber
\\
&&=\Gamma(\mu)e^{-a}U\left(\frac{1}{2}+\mu-\kappa+\lambda,
1+2\lambda,a\right),
\ \ [\Re{\mu}>0, \ \Re{a}>0],
\end{eqnarray}
which is equivalent to \cite{gradshteyn}
\begin{eqnarray*}
\int_{1}^{\infty}
e^{-a y/2}(y-1)^{\mu-1}y^{\kappa-\mu-1}W_{\kappa,\lambda}(a y)
dy=\Gamma(\mu)e^{-a/2}
W_{\kappa-\mu,\ \lambda}(a), 
\ \ [\Re{\mu}>0, \ \Re{a}>0].
\end{eqnarray*}
%
%
%
%
%BIBLIOGRAFIA
%
%


\begin{thebibliography}{99}
%
%
\bibitem{eu} Figueiredo B D B {\it 2002} On some solutions to 
generalized  spheroidal wave equations and applications {\it J. Phys. A: Math. Gen.} 
{\bf 35} 2877; {\bf 35} 4799 (corrigendum).
%
%
\bibitem{wilson}Wilson A H 1928 A generalized spheroidal wave equation 
{\it Proc. Roy. Soc. London} {\bf A118} 617.
%
%
\bibitem{ronveaux} Ronveaux A (editor) 1995 {\it Heun's differential 
equations} (Oxford University Press).
%
%
\bibitem{leaver1}Leaver E W 1986 Solutions to a generalized spheroidal 
wave equation: Teukolsky equations in general relativity, and the 
two-center problem in molecular quantum mechanics {\it J. Math. Phys.} 
{\bf 27} 1238.
%
%
\bibitem{schmidt} Schmidt D and Wolf G 1995 Double confluent Heun equation, 
Part C of \cite{ronveaux}.
%
\bibitem{slavyanov} Slavyanov S Yu and Lay W 2000 {\it Special functions: 
a unified theory based on singularities} (Oxford University Press).
%
\bibitem{decarreaux1} Decarreau A, Dumont-Lepage M C, Maroni P, Robert A 
and Ronveaux A 1978 Formes canoniques des \'equations confluentes de 
l'\'equation de Heun {\it Ann. Soc. Sci. Brux.} {\bf T92(I-II)} 53.
%
\bibitem{decarreaux2}Decarreau A, Maroni P and Robert A 1978 Sur les 
\'equations confluentes de l'\'equation de Heun {\it Ann. Soc. Sci. Brux.} 
{\bf T92(III)} 151.
%
\bibitem{ince} Ince E L 1926 {\it Ordinary differential equations} 
(New York: Dover Publications).
%
\bibitem{olver}Olver F W J 1974 {\it Asymptotics and special functions}
(New York: Academic Press).
%
\bibitem{gautschi}Gautschi W 1967 Computational aspects of three-term 
recurrence relations {\it SIAM Review} {\bf 9} 24. 
%
\bibitem{abramowitz} Abramowitz M and Stegun I A (eds.) 1965 
{\it Handbook of Mathematical Functions} (New York: Dover).
%
\bibitem{erdelyi1}Erd\'elyi A {\it et al.} 1953 {\it Higher Transcendental 
Functions} Vol. 1 (McGraw-Hill, New York).
%
\bibitem{arscott} Arscott F M 1964 {\it Periodic Differential 
Equations} (Oxford: Pergamon Press).
%
\bibitem{lemieux}Lemieux A and Bose A K {\it 1969}  
Construction de potientels pour lesquels l'\'equation de Schr\"{o}dinger 
est soluble {\it Ann. Inst. Henri Poincar\'e} 
{\bf 10} 259.
%
%
\bibitem{arscott1} Arscott F M  1967 The Whittaker-Hill 
equation and the wave equation in paraboloidal co-ordinate 
{\it Proc. Roy. Soc. Edinburg} {\bf A67} 265.
%
%
\bibitem{zaslavskii}Zaslavskii O B and Ulyanov V V {\it 1984} New classes
of exact solutions of the Schr\"{o}dinger equation and the
potential-field description of spin systems {\it Sov. Phys. JETP} {\bf
60} 991; Ulyanov V V and Zaslavskii O B 1992 New methods in the theory
of quantum spin systems {\it Phys. Rep.} {\bf 216} 179.
%
%
\bibitem{usheveridze}Ushveridze A G {\it 1989} Quasi-exactly solvable models in 
quantum mechanics {\it Sov. J. Part. Nucl.} {\bf 20} 504; 1994 
{\it Quasi-exactly solvable models in quantum mechanics} (IOP Publishing). 
%
%
\bibitem{gonz}Gonz\'alez-L\'opez A, Kamran N and Olver P J 1993 
Normalizability of one-dimensional quasi-exactly solvable Schr\"{o}dinger
operators {\it Commun. Math. Phys.} {\bf 153} 117.
%
\bibitem{buhring}B\"{u}hring W 1974 Schr\"{o}dinger equation 
with inverse fourth-power potential, a differential equation with two 
irregular singular points {\it J. Math. Phys.} {\bf 15} 1451.
%
\bibitem{bose}Bose S K and Gupta N 1998  
Exact solution of nonrelativistic Schr\"{o}dinger equation for certain
central physical potentials {\it Il Nuovo Cimento} {\bf 113B} 299.
%
%
\bibitem{dong}Dong S H and Ma Z Q 1998 Exact solutions to the 
Schr\"{o}dinger equation for the potential $V(r)=ar^2+br^{-4}+cr^{-6}$
in two dimensions {\it J. Phys. A: Math.Gen.} {\bf 31} 9855.
%
\bibitem{gonul}G\"{o}n\"{u}l B, \"{O}zer O, Ko\c{c}ak M, Tutcu D and
Can\c{c}elik Y 2001 Supersymmetry and the relationship between a class
of singular potentials in arbitrary dimensions {\it J. Phys. A:
Math. Gen.} {\bf 34}, 8271.
%
%
\bibitem{bagchi}Bagchi B and Ganguly A 2003 A unified treatment 
of exactly solvable and quasi-exactly solvable quantum potentials 
{\it J. Phys. A: Math. Gen.} {\bf 36}, L161-L167.
%
%
\bibitem{prudnikov}Prudnikov A P, Brychkov Yu A and Marichev O I 1990
Integrals and series, v.3: More special functions (Gordon and Breach Science
Publishers, New York).
%
\bibitem{gradshteyn}Gradshteyn I S and Ryzhik I M 1994 
{\it Table of integrals, series and products} (Academic Press, New York).
%
%
\end{thebibliography}
\end{document}